\documentclass[aip,jcp]{revtex4-1}
\usepackage{graphicx}
\usepackage{amsmath}
\usepackage{subfigure}
\usepackage{braket}
\usepackage{color}
\usepackage[dvipsnames]{xcolor}
\usepackage{epstopdf}
\usepackage[normalem]{ulem}
\usepackage{ulem}
\definecolor{ao(english)}{rgb}{0.0, 0.42, 0.24}

\newcommand{\MC}[1]{\textcolor{black}{{#1}}}

\newcommand{\RX}[1]{\textcolor{black}{{#1}}}
\newcommand{\RXN}[1]{\textcolor{black}{{#1}}}

\newcommand{\XY}[1]{\textcolor{purple}{{#1}}}
\usepackage[utf8]{inputenc}
\usepackage[T1]{fontenc}
\usepackage{mathptmx}
\usepackage{etoolbox}
\makeatletter
\def\@email#1#2{%
 \endgroup
 \patchcmd{\titleblock@produce}
  {\frontmatter@RRAPformat}
  {\frontmatter@RRAPformat{\produce@RRAP{*#1\href{mailto:#2}{#2}}}\frontmatter@RRAPformat}
  {}{}
}%

\begin{document}

\title{Structural and Dynamic Properties of Solvated Hydroxide and Hydronium Ions in Water from {\it Ab Initio} Modeling}

\author{Renxi Liu}
\affiliation{HEDPS, CAPT, College of Engineering, Peking University, Beijing, 100871, China}
\affiliation{Academy for Advanced Interdisciplinary Studies, Peking University, Beijing, 90871, China}
\author{Chunyi Zhang}
\affiliation{Department of Physics, Temple University, Philadelphia, PA 19122, USA}
\author{Xinyuan Liang}
\affiliation{HEDPS, CAPT, College of Engineering, Peking University, Beijing, 100871, China}
\affiliation{Academy for Advanced Interdisciplinary Studies, Peking University, Beijing, 90871, China}
\author{Jianchuan Liu}
\affiliation{HEDPS, CAPT, College of Engineering, Peking University, Beijing, 100871, China}
\author{Xifan Wu}
\affiliation{Department of Physics, Temple University, Philadelphia, PA 19122, USA}
\affiliation{Institute for Computational Molecular Science, Temple University, Philadelphia, PA 19122, USA}
\author{Mohan Chen}
\email{mohanchen@pku.edu.cn}
\affiliation{HEDPS, CAPT, College of Engineering, Peking University, Beijing, 100871, China}
\affiliation{Academy for Advanced Interdisciplinary Studies, Peking University, Beijing, 90871, China}

\date{\today}

\begin{abstract}
Predicting the asymmetric structure and dynamics of solvated hydroxide and hydronium in water has been a challenging task from {\it ab initio} molecular dynamics (AIMD). The difficulty mainly comes from a lack of accurate and efficient exchange-correlation functional in elucidating the amphiphilic nature and the ubiquitous proton transfer behaviors of the two ions.
By adopting the strongly-constrained and appropriately normed (SCAN) meta-GGA functional in AIMD simulations,
we systematically examine the amphiphilic properties,
the solvation structures, the electronic structures, and the dynamic properties
of the two water ions. In particular, we compare these results to those predicted by the PBE0-TS functional,
which is an accurate yet computationally more expensive exchange-correlation functional.
We demonstrate that the general-purpose SCAN functional provides
a reliable choice in describing the two water ions.
Specifically, in the SCAN picture of water ions,
the appearance of the fourth and fifth hydrogen bonds near hydroxide stabilizes the pot-like shape solvation structure and suppresses the structural diffusion, while the hydronium stably donates three hydrogen bonds to its neighbors. 
\RX{We apply a detailed analysis of the proton transfer mechanism of the two ions and find the two ions exhibit substantially different proton transfer patterns.}
\MC{Our AIMD simulations indicate} hydroxide diffuses slower than hydronium in water, which is consistent with the experiments.
\end{abstract}

\maketitle

\section{Introduction}

Hydroxide and hydronium are two ubiquitous ions
behind all acid-base reactions in water.
A variety of chemical reactions in an aqueous environment are influenced by the existence of the two ions. Therefore, understanding the structure and dynamics of both ions is important for the studying of water and various related research areas.\cite{16CR-Agmon}
In addition, the solvated hydroxide $\rm OH^{-}$(aq) and hydronium $\rm H_{3}O^{+}$(aq) ions in water exhibit intriguing properties.
For example, as explained by the Grotthuss mechanism,~\cite{Grotthuss}
the diffusion of hydroxide (hydronium) ion in an aqueous environment can be viewed as
an excessive proton hole (proton) that hops among water molecules through continuous breakage and formation of hydrogen bonds (HBs) in water.
This so-called proton transfer (PT) process leads to abnormally high diffusivities of the two ions,
which have important implications in a wide variety of biological, environmental
and industrial processes.~\cite{96CM-Kreuer, 06BBAB-Samuel, 06BBAB-Collin}
However, the structural similarity of the two ions does not bring to similar diffusivity,
rather, hydroxide diffuses twice as slow as hydronium does.~\cite{83JCS-Halle, 83RSC-Halle, 90N-Weingartner, 10JPCB-Sluyters}
In fact, the PT process occurs on a relatively short timescale of picoseconds and is
largely influenced by the H-bond network in water. 
%
%
\MC{Until now, experimental evidences of PT come from indirect measurements such as nuclear magnetic resonance,~\cite{61JCP-Meiboom} from which Agmon deduced the modern version of Grotthuss mechanism for hydronium.~\cite{95CPL-Agmon}
}
\RXN{Two-dimensional infrared spectroscopy captures the response of some characteristic vibration modes to perturbations of certain frequency,
providing some indirect evidences for PT in both ions as well.~\cite{18JPCB-Carpenter, 18NC-Founier, 09PNAS-Roberts, 11JPCA-Roberts, 15S-Thamer, 18JCP-Napoli, 19ACS-Yuan}}
Instead, {\it ab initio} molecular dynamics (AIMD) can access the timescale of PT events and therefore
plays an important role in explaining the PT events.~\cite{96N-Luzar, 99N-Marx} 
\MC{In addition, empirical force fields such as ReaxFF force field~\cite{17JPCB-Zhang} and multiscale empirical valence bond (MS-EVB) model~\cite{08JPCB-Wu, 16JPCB-Biswas} can be used to gain further microscopic insights into aqueous hydronium~\cite{08JPCB-Omer, 15JCP-Tse, 21JACS-Calio} and hydroxide~\cite{15JACS-Tse, 17JPCB-Zhang}.
Note that the accuracy of the empirical methods largely depends on tuned parameters, some of which were obtained from accurate AIMD simulations.~\cite{08JPCB-Omer, 21JACS-Calio} In this work, we focus on discussing the accuracy and efficiency of AIMD methods \RX{in describing the solvated hydroxide and hydronium ions.}
}

The prerequisite of simulating the above two ions in water is an accurate description of bulk water from AIMD.
In the last few decades, AIMD simulations based on the density functional theory (DFT)~\cite{64PR-Hohenberg,65PR-Kohn}
have been widely utilized to characterize bulk water because the accuracy and efficiency are reasonably balanced.~\cite{94JPCM-Tuckerman, 06ACR-Tuckerman, 18NC-Chen, 11PNAS-Hassanali, 13PNAS-Hassanali, 19AR-Sakti}
However, modelling of bulk water is non-trivial because several competing physical effects are present.
Specifically, bulk water at ambient conditions is comprised of water molecules embedded within
a continuously fluctuating H-bond network,
which originates from a delicate balance among various physical interactions including
covalent bonds, H-bonds, and van der Waals interactions.~\cite{15JPCL-Gaiduk, 17PNAS-Chen, 14JCP-Distasio}
Importantly, the balance between intra-molecular and inter-molecular interactions in liquid water largely depends on the choice of exchange-correlation (XC) functionals utilized in DFT.
Early AIMD simulations of water~\cite{92CPL-Laasonen, 93CPL-Laasonen}
based on the LDA or GGA XC functionals predicted an over-structured H-bond network of water,
resulting in sluggish and rigid water with a low density.~\cite{16JCP-Gillan, 15JPCL-Gaiduk}
Later, adding the van der Waals interactions upon GGA functionals has become a common practice to yield a better description of structural and dynamic properties of water, as well as \MC{to correct the water density that is systematically underestimated by GGA functionals} at ambient conditions,
\cite{15JPCL-Gaiduk, 09JPCB-Schmidt, 11JCP-Wang, 15JCP-Miceli, 17PNAS-Chen, 14JCP-Distasio}
\RXN{although the addition of dispersion correction and the corrected AIMD results could be sensitive to the choice of basis sets.~\cite{12JCP-Ma}}
On \MC{the} one hand, the van der Waals interactions increase
the non-directional attraction forces between water molecules, which compensate for the underestimated water density.~\cite{17PNAS-Chen}
On the other hand, the PBE0 hybrid functional~\cite{perdew1996rationale, adamo1999toward} mitigates the
self-interaction error and better accounts for the molecular polarizability in liquid water simulations,~\cite{14JCP-Distasio}
resulting in weakened HBs between water molecules that are closer to experiment.
The above two corrections draw the accuracy of simulating water towards experimental data.

Simulating $\rm H_{3}O^{+}$(aq) and $\rm OH^{-}$(aq) ions in water
is more challenging than bulk water.
\RXN{First, the PT mechanism of hydronium has been intensively studied.~\cite{94JPCM-Tuckerman, 95JCP-Tuckerman, 15JCP-Tse, 18NC-Chen, 09L-Berkelbach, 05JCP-Izvekov, 07L-Chandra}
It is predicted that the hydronium ion transfers through the HB network by quickly interchanging between Zundel ($\mathrm{H_{5}O_{2}^{+}}$)~\cite{68ZFPC-Zundel} and Eigen ($\mathrm{H_{3}O^{+}(H_2O)_{3}}$)~\cite{58PRSL-Eigen} complex.
In particular, the BLYP XC functional predicted a rearrangement of the HB network around hydronium in tens of femtoseconds before PT, including the breakage of a HB donated to the first-shell water molecule that tends to accept the excess proton, and the formation of a weak HB accepted by the hydronium.~\cite{09L-Berkelbach}
Later study suggested that the formation of a weak HB on the O end of hydronium could be a sign of PT burst period.~\cite{15JCP-Tse, 16JPCB-Biswas}}
%
Different GGA functionals such as PW91~\cite{98-Burke}, HCTC~\cite{00JCP-Boese}, and BLYP~\cite{88A-Becke, 88B-Lee, 06ACR-Tuckerman,19AR-Sakti}
gave rise to dramatically different solvation structures and dynamics of solvated hydroxide.
Therefore, the difficulty is particularly prominent in modeling hydroxide
because it has more than one solvation structure.
For example, the hydroxide ion described by PW91 favors a 3-fold coordination structure (tetrahedral structure),
while those described by HCTH and BLYP favor a 4-fold coordination structure (hyper-coordination structure).~\cite{06ACR-Tuckerman}
Previous studies proposed that the hyper-coordination structure
strongly suppresses the PT events. \RXN{In detail, a hydroxide described by BLYP allows structural diffusion by transforming a hyper-coordination structure to a three-coordination structure to occur more easily than the one described by the HCTH functional. }~\cite{02N-Tuckerman, 06ACR-Tuckerman, 11CPL-Ma}
As a result, the two distinct H-bond structures lead to distinct diffusivities
of the hydroxide ion.
\RXN{It is also proposed with the BLYP functional that the percentage of hyper-coordinated structure increases as temperature lowers, resulting in a low PT rate and a slow reorientation process at low temperatures.~\cite{11CPL-Ma}}
Although the BLYP functional predicts reasonable diffusion coefficients
of hydroxide, it substantially underestimates the density of liquid water (\RXN{BLYP functional with different bases predicts density ranging from $\rm 0.75~g/cm^{3}$ to $\rm 0.92~g/cm^3$.~\cite{05CPC-Ma, 09JPCB-Schmidt, 12JCP-Ma}}),
implying that the BLYP functional is not adequate to describe the sophisticated H-bond network of liquid water.
Importantly, the BLYP functional predicted
that both hydronium and hydroxide ions prefer to experience bursts of PT events
than single PT events within a timescale of water wire compression.~\cite{11PNAS-Hassanali,13PNAS-Hassanali}
These ubiquitous concerted PT behaviors of the two ions are highly correlated in time, which adds
a new twist to the already sophisticated story but does not provide enough evidence to
explain the slower diffusivity of $\rm OH^{-}$(aq) than that of $\rm H_{3}O^{+}$(aq).

The concerted PT picture is updated via considering both exact exchange
and van der Waals interactions in AIMD simulations.~\cite{18NC-Chen}.
By utilizing the hybrid functional PBE0~\cite{perdew1996rationale} with
the self-consistent Tkatchenko-Scheffler (TS) functional in AIMD simulations,~\cite{09vdW-TS,14JCP-Distasio}
the so-called PBE0-TS exchange-correlation functional mitigates the spurious self-interaction and includes both intermediate- and long-range van der Waals interactions. This leads to a better description of the H-bond network of liquid water,
where structural and dynamic properties of $\rm OH^{-}$(aq) and $\rm H_{3}O^{+}$(aq) are
substantially improved towards the experimental data.
For example, while previous AIMD simulations utilizing GGA functionals
suggested that the hydroxide has a planar solvation structure,~\cite{06ACR-Tuckerman, 06CPC-Marx}
the PBE0-TS predicted a pot-like solvation structure of hydroxide that agrees with the
picture from neutron scattering data.~\cite{03JCP-Botti, 18NC-Chen}
In addition, unlike BLYP, the PBE0-TS functional largely stabilized the 4-fold solvation structure of $\rm OH^{-}$(aq),
which inhibits the concerted PT behaviors of hydroxide.
Instead, the solvated hydroxide ion described by PBE0-TS exhibits a stepwise motion.
Meanwhile, $\rm H_{3}O^{+}$(aq) still preserves
the concerted PT behavior within the description of PBE0-TS. Consequently, PBE0-TS predicts
a slower diffusivity of hydroxide than hydronium,~\cite{18NC-Chen} which agrees well with the experiments.~\cite{83RSC-Halle, 90N-Weingartner, 10JPCB-Sluyters}

Unfortunately, despite the apparent merits of utilizing linear-scaling PBE0-TS~\cite{20JCTC-Ko}
in studying liquid water and ions, the prefactor of the linear-scaling method is still large and the computational costs are extremely high.
In this regard, the community awaits an accurate yet efficient XC functional that can be utilized to study
not only liquid water but also $\rm OH^{-}$(aq) and $\rm H_{3}O^{+}$(aq) ions.
Notably, the recently proposed SCAN functional,~\cite{15L-Sun}
fulfilling all 17 known exact requirements on the semilocal XC functional, can achieve similar accuracy as PBE0-TS does in describing the structural and dynamic properties of liquid water at ambient conditions, as well as the water density.\cite{17PNAS-Chen}
In particular, the computational cost of SCAN is comparable to GGA functionals but about an order of magnitude smaller than the PBE0-TS functional for studying condensed phases.\cite{20JPCL-Furness}
%
%
Compared to GGA functionals,
SCAN predicts stronger covalent bonds within water molecules and less negatively charged local environment around the oxygen atom.
As a result, the directional HBs are weakened.
In addition, the intermediate-ranged van der Waals interaction in SCAN provides attractive forces
among water molecules, which leads to the movements of the second-shell water molecules towards the
non-H-bonded interstitial area, creating a more disordered and denser structure
on the intermediate range.
The above changes provided by SCAN results in a weakened H-bond network of liquid water that is closer to the experimental data.~\cite{NIST-book, CRC-book}
As a result, SCAN provides better structural and dynamic properties of liquid water than PBE.~\cite{17PNAS-Chen}
This is considered to be the main reason why SCAN,
rather than PBE and other GGA functionals, predicts the correct quantitative relation
between densities of water and ice under ambient conditions.~\cite{17PNAS-Chen}
Furthermore, SCAN also exhibits high precision in modeling water clusters, gas and ice phase water,
which outperforms GGA functionals.~\cite{16NC-Sun, 18JCP-Zheng, 19B-Xu, 20PNAS-Sharkas, 21B-Xu}

Nevertheless, to the best of our knowledge,
the effects of the SCAN functional in describing the structure and
dynamics of water ions have not been thoroughly explored yet.
In this work, we have performed AIMD simulations of $\rm OH^{-}$(aq) and $\rm H_{3}O^{+}$(aq) ions
using the SCAN exchange-correlation functional.
Detailed analysis was performed on these AIMD trajectories to evaluate the amphiphilic properties,
the solvation structures, the electronic structures, and the dynamic properties.
In particular, the SCAN results are systematically compared to the PBE0-TS results,~\cite{18NC-Chen}
which serve as a valid benchmark to evaluate the performances of SCAN for $\rm OH^{-}$(aq) and $\rm H_{3}O^{+}$(aq) ions.
The paper is organized as follows, Section II introduces the computational details. The
AIMD results are shown and discussed in Section III, and the concluding remarks are presented in
Section IV.

\section{Computational Details}
We have performed AIMD simulations by using the Car-Parrinello molecular dynamics method~\cite{85L-CPMD}
implemented in the Quantum ESPRESSO package.~\cite{codeQE}
We used 63 water molecules and a hydroxide or hydronium ion
in a periodic cubic cell with a length of 12.4447~\AA.
A kinetic energy cutoff of 85 Ry was adopted.
We used the meta-GGA exchange-correlation functional in the form of SCAN~\cite{15L-Sun}
and the Hamann-Schl$\ddot{u}$ter-Chiang-Vanderbilt pseudopotentials
generated with the PBE functional.~\cite{79L-Hamann}
For the AIMD simulations, we adopted the Nos\'{e}-Hoover chain thermostats with a chain length
of 4 for each ion to control the temperature of 330 K in the NVT ensemble.~\cite{84JCP-Nose,85A-Hoover,92JCP-Martyna}
The mass of hydrogen atom was set to 2.0141 (mass of deuterium).
A fictitious electron mass of 100 a.u. was used along with a mass cut-off of 25 a.u.
The AIMD trajectory lengths were 124.4 and 176.9 ps for the hydroxide and hydronium simulations, respectively. A time step of 2.0 a.u was adopted.
AIMD results utilizing the PBE0-TS XC functional referenced
from literature~\cite{18NC-Chen}
were also shown, which serve as a baseline to
elucidate the effects of the SCAN functional in predicting properties of
hydronium and hydroxide in an aqueous environment.

A standard hydrogen bond criterion was adopted\cite{96N-Luzar}.
Specifically, two neighboring water molecules are defined to be H-bonded when
the O-O distance is less than 3.5 $\rm \AA$,
and the O-O-H angle is less than $30^{\circ}$.
\MC{A covalent bond is defined between an oxygen atom and a hydrogen atom when the distance between their nuclei is less than 1.24 $\rm \AA$.
%
%
A hydroxide is defined as a molecule with only one H atom forming a covalent bond with the O atom,
while a hydronium is defined as a molecule with three H atoms forming covalent bonds with the O atom.
}

\begin{figure}
  \includegraphics[width=8.5cm]{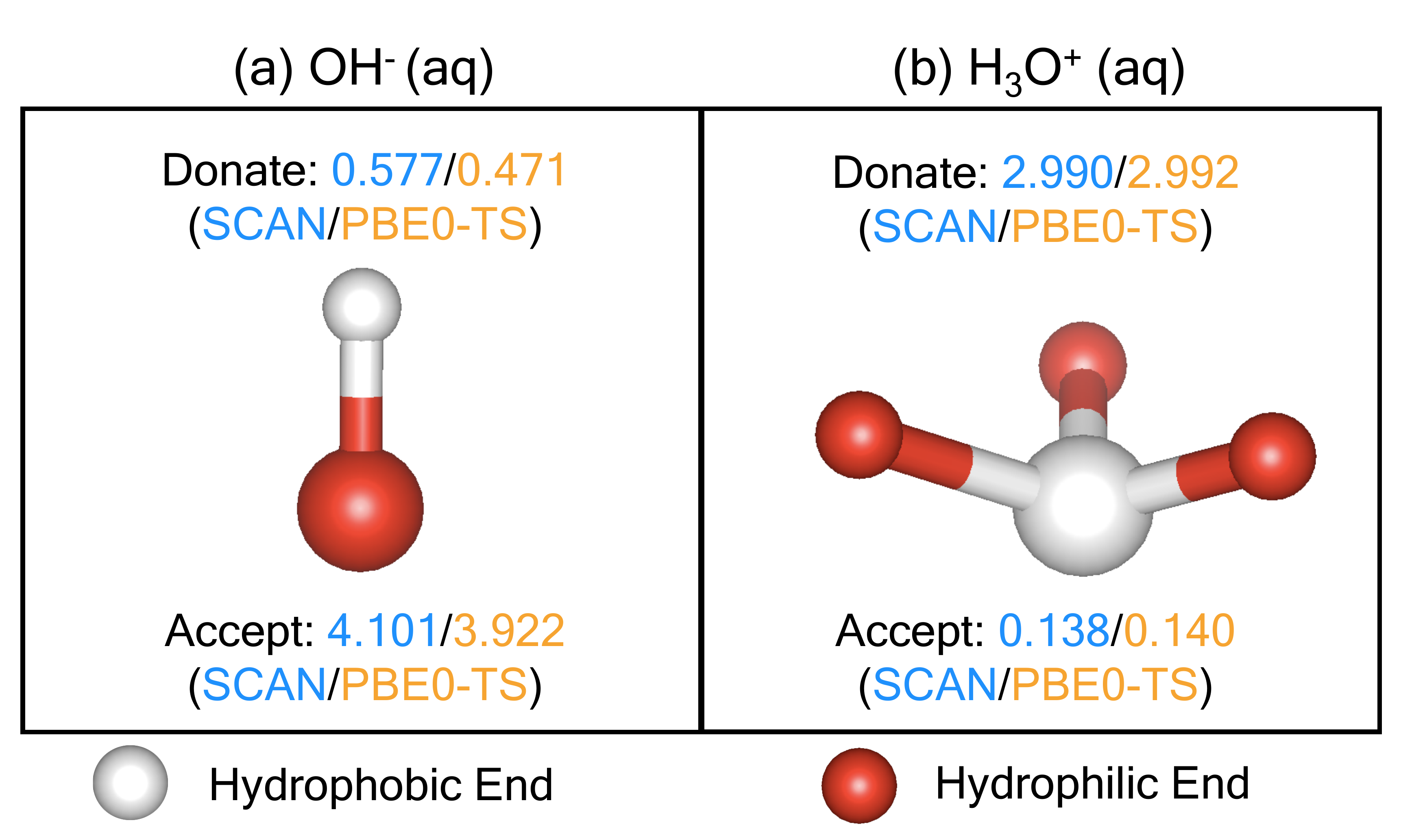}
  \caption{
  (Color online) Hydrophobic (gray) and hydrophilic (blue) ends in solvated (a) $\mathrm{OH^{-}}$(aq) and (b) $\mathrm{H_{3}O^{+}}$(aq) ions.
  Average number of accepted and donated hydrogen-bonds (HBs) for both ions is listed. Two exchange-correlation functionals, i.e., SCAN and PBE0-TS, are used.
  }
  \label{amph}
\end{figure}

\begin{figure}
  \includegraphics[width=8.5cm]{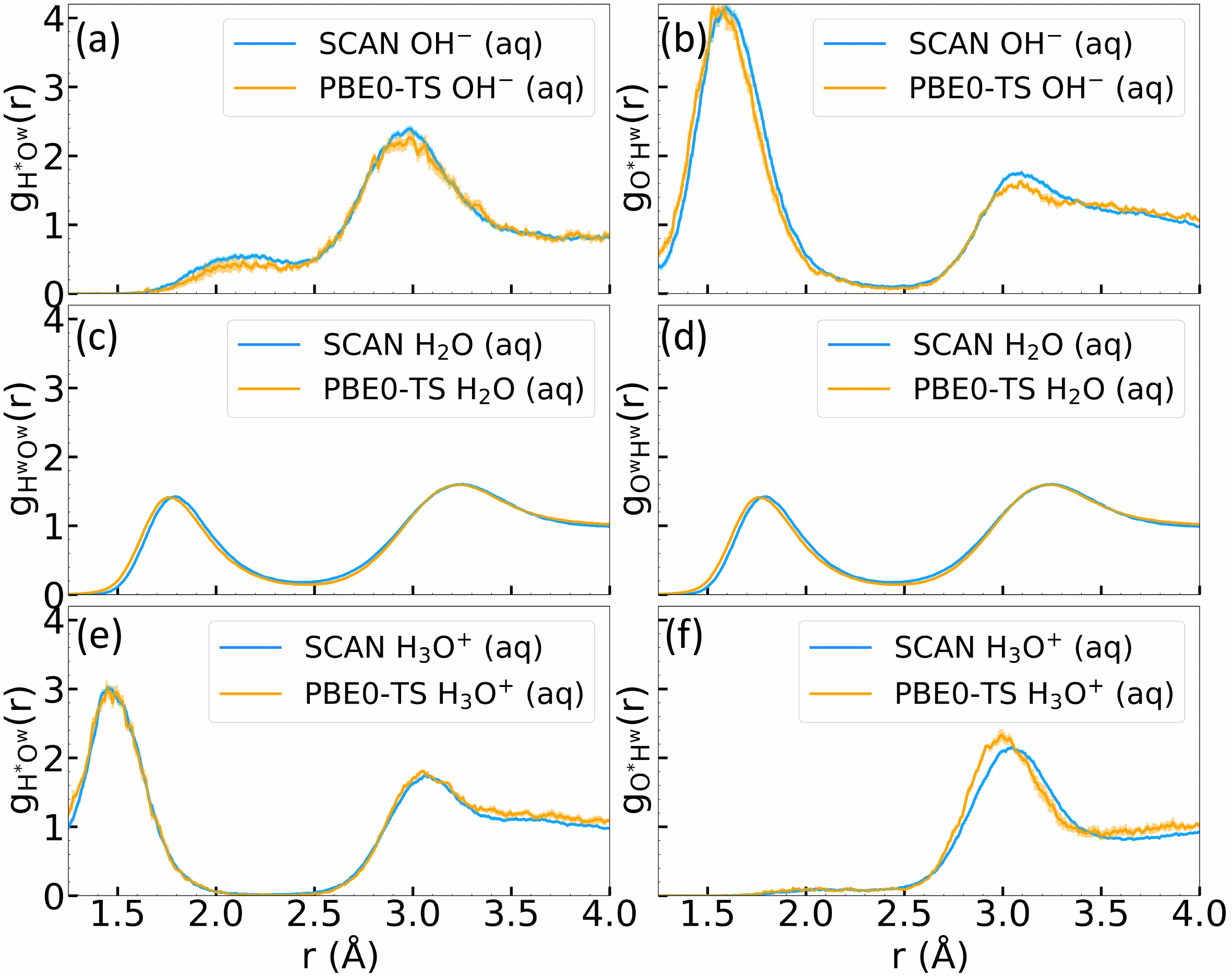}
  \caption{
(Color online) Radial distribution functions $\rm g(r)$ as extracted from
  SCAN (blue) and PBE0-TS (orange) trajectories. Superscripts $\rm *$ and $\rm w$ depict atoms in ions
  and water molecules, respectively. (a), (c) and (e) respectively show the distribution of O atoms
  around H atoms in the $\rm OH^{-}$(aq), $\rm H_{2}O$ and $\rm H_{3}O^{+}$(aq) molecules,
  while those in (b), (d) and (f) display
  the distribution of H atoms around O atom in the $\rm OH^{-}$(aq),
  $\rm H_{2}O$ and $\rm H_{3}O^{+}$(aq) molecules, respectively. \MC{Error bars are illustrated for each curve in light colors. }} \label{pdf}
\end{figure}

\section{Results and Discussion}

\subsection{Amphiphilic Properties}


Both $\rm OH^{-}$(aq) and $\rm H_{3}O^{+}$(aq) ions are amphiphilic in nature as compared to the water molecules. 
Fig.~\ref{amph} shows the hydrophobicity and hydrophilicity of different ends in both ions, as well as the number of HBs for each end.
The oxygen site of $\rm OH^{-}$(aq) is hydrophilic and the hydrogen site is hydrophobic; on the contrary, the oxygen site of $\rm H_{3}O^{+}$(aq) is hydrophobic while the hydrogen site is hydrophilic.
Fig.~\ref{pdf} illustrates the radial distribution functions $\rm g(r)$ of solvated hydroxide and hydronium ions,
as well as those of the ambient liquid water from AIMD simulations using both SCAN and PBE0-TS XC functionals. Note that the superscripts $\rm *$ and $\rm w$ denote an atom from an ion and a water molecule, respectively.
\MC{The error bars of each curve shown in Fig.~\ref{pdf} represent the standard errors obtained by dividing the MD trajectory into several segments with the length of each segment being 2.419 ps (10,000 snapshots); these standard errors of sampling are relatively small compared to the amplitude of the radial distribution functions. Therefore, the errors do not influence the discussion below.
}
On \MC{the} one hand,
the two ions are hydrophobic, as respectively illustrated in Figs.~\ref{pdf} (a) and (f),
$\rm g_{H^{*}O^{w}} (r)$ of hydroxide and $\rm g_{O^{*}H^{w}} (r)$ of hydronium both exhibit the reduced HBs
of the first solvation shell as compared to ambient liquid water. Specifically, the hydroxide (hydronium) ion shows a strong repulsion against water molecules on its hydrogen (oxygen) site by largely reducing the donated (accepted) HBs with neighboring atoms.
On the other hand,
the two ions are hydrophilic, which can be seen in Figs.~\ref{pdf}(b) and (e)
as the first peak of $\rm g(r)$ shifts to the left when compared to its counterpart in pure liquid water.
In detail, for the first high peak of $\rm g_{O^{*}H^{w}} (r)$ shown in Fig.~\ref{pdf}(b),
the oxygen site of the hydroxide ion attracts multiple neighboring water molecules via forming HBs,
while the first high peak of $\rm g_{H^{*}O^{w}} (r)$ shown in Fig.~\ref{pdf}(e) implies that
the hydrogen sites of the hydronium stably donate three protons to neighboring water molecules and form HBs.
As will be explained below, the amphiphilic nature of the two ions leads to their unique structural and dynamic properties.


Unlike some GGA functionals that yield qualitatively different radial distribution functions for hydroxide,~\cite{06ACR-Tuckerman}
the SCAN functional yields a $\rm g(r)$ similar to that predicted by PBE0-TS, as shown in Fig.~\ref{pdf}.
Regarding the computational costs, SCAN takes the advantage of being much more efficient than the hybrid functional.~\cite{17PNAS-Chen}
Therefore, it is valuable to compare properties of hydroxide and hydronium
in detail as obtained from the two functionals and we have the following findings.
First, the hydrophobic sites of the two ions are investigated.
We observe a less hydrophobic end in hydroxide (hydrogen site) as compared to that in hydronium (oxygen site),
because the first and the second peaks of $\rm g_{H^{*}O^{w}}(r)$ for hydroxide in Fig.~\ref{pdf}(a) are considerably
higher than those in the $\rm g_{O^{*}H^{w}}(r)$ for hydronium shown in Fig.~\ref{pdf}(f), suggesting more water molecules locate around the hydrogen site of hydroxide than the oxygen site of hydronium.
When comparing the results from SCAN to those from PBE0-TS, the hydrophobic site of hydroxide, i.e.,
the hydrogen of hydroxide, becomes less hydrophobic by donating slightly more HBs to its neighbors.
This is evidenced as SCAN yields a higher first peak of $\rm g_{H^{*}O^{w}}(r)$ than PBE0-TS in Fig.~\ref{pdf}(a).
The result is also consistent with the HB data shown in Fig.~\ref{amph}, where the averaged number of donating HBs in hydroxide is 0.577 and 0.471 from SCAN and PBE0-TS, respectively.
On the contrary, the hydrophobic site of hydronium, i.e., the oxygen site, becomes more hydrophobic by adopting the SCAN functional, which is evidenced by a lower first peak of $\rm g_{O^{*}H^{w}} (r)$ located around 3~\AA~in Fig.~\ref{pdf}(f).
Moreover, the peak locates further away from the central atom than the one from PBE0-TS, suggesting that it would be more difficult for hydronium to accept HBs in SCAN (0.138) than in PBE0-TS (0.140), as listed in Fig.~\ref{amph}.

Second, we focus on the hydrophilic sites of the two ions, where multiple HBs form between the ion and water molecules.
Notably, the symmetric picture for the twin topological defects, i.e., hydroxide and hydronium ions,
is invalid from both XC functionals if we inspect the coordination number of the two ions.
Specifically, the height of the first peak at the hydrophilic end of hydroxide (Fig.~\ref{pdf}(b))
is higher than that of hydronium (Fig.~\ref{pdf}(e)),
indicating a larger coordination number.
For example, according to the HB analysis of the SCAN trajectory in Fig.~\ref{amph}, hydroxide accepts 4.101 HBs on average, which is larger than hydronium that donates 2.990 H-bonds.
Meanwhile, the oxygen site of hydroxide described by PBE0-TS donates 3.922 HBs, which exhibits less hydrophilic property as compared to the one by SCAN.
Furthermore, we observe more non-H-bonded water molecules in the second shell of hydroxide from the SCAN functional, which can be seen by a higher second peak of $\rm g_{O^{*}H^{w}} (r)$ in Fig.~\ref{pdf}(b).
%
%
The above structural features of hydroxide imply that SCAN gathers more H-bonded water molecules
around both ends of hydroxide than PBE0-TS.
%

\begin{figure}
  \includegraphics[width=8.5cm]{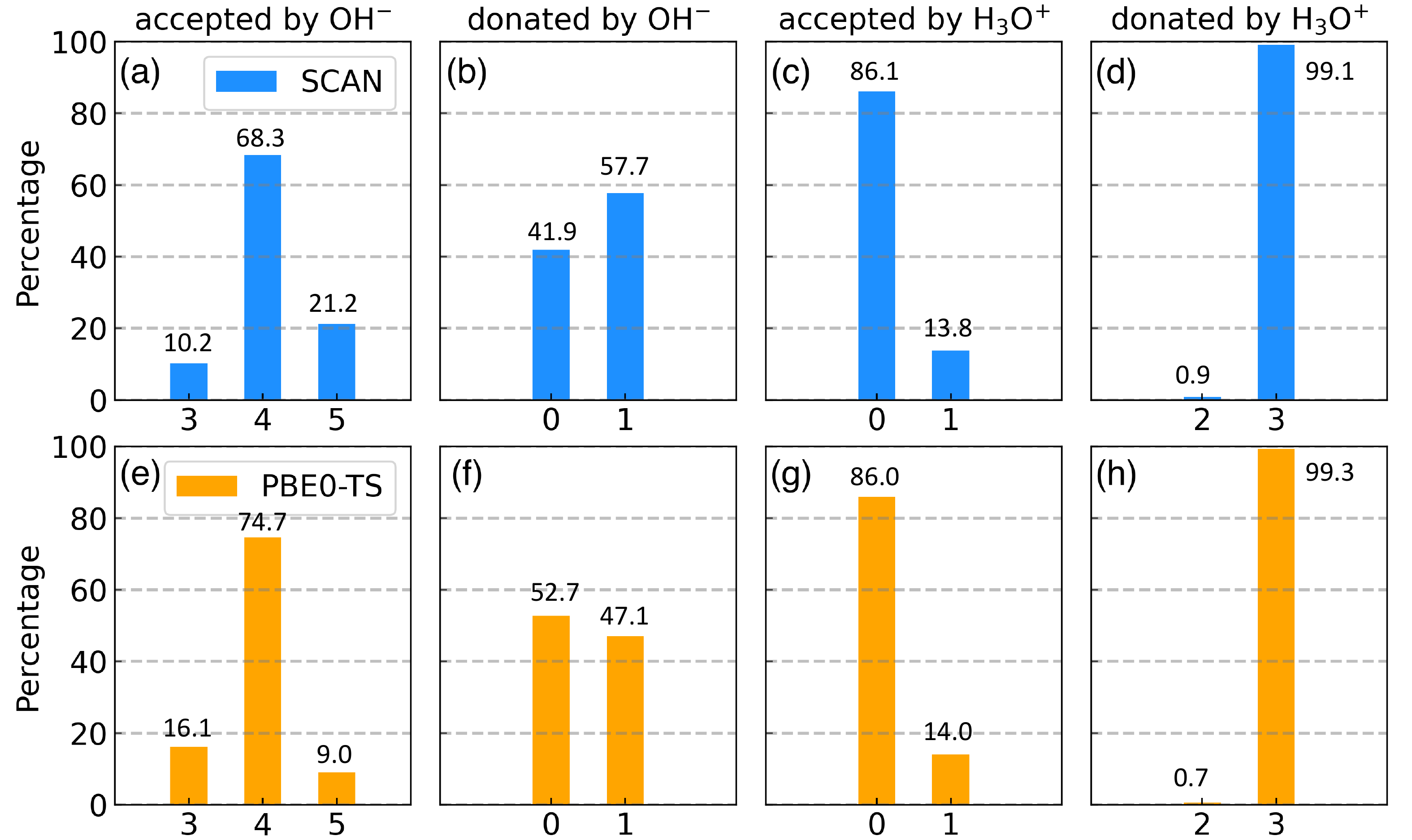}
  \caption{(Color online) Percentages of H-bonds
  for $\rm OH^{-}$(aq) and $\rm H_{3}O^{+}$(aq) ions.
  The first row (a-d) presents results from the SCAN functional (blue),
  while the second row (e-h) displays those from the PBE0-TS functional (orange).
  }  \label{HB}
\end{figure}

The amphiphilic nature of the two ions can be further analyzed by the HB statistics of the hydroxide and hydronium ions.
As found by previous studies\cite{06ACR-Tuckerman,18NC-Chen}, the hydronium cation stably donates three HBs but the hydroxide ion has two H-bonded solvation structures, i.e., a nearly tetrahedral structure with three acceptor HBs and a hyper-coordinated structure with four or more acceptor HBs. While the PBE functional predicts more tetrahedral-like structures of hydroxide,
the inclusion of van der Waals interactions and self-interaction correction
helps to soften and stabilize the hypercoordination of hydroxide.~\cite{18NC-Chen}
Importantly, both SCAN and PBE0-TS functionals include the van der Waals interactions
and predict a majority of hydroxide ions that accept four or more HBs to form the so-called hyper-coordinated structure,
which are shown in Figs.~\ref{HB}(a) and (e).
For instance, 68.3\% and 74.7\% of hydroxide ions with four accepted HBs are found in the SCAN and PBE0-TS trajectories, respectively.
Interestingly, the SCAN functional predicts substantially more hydroxide ions that accept 5 HBs
(21.2\%) than the PBE0-TS functional (9.0\%).
Additionally, SCAN and PBE0-TS give rise
to 10.2\% and 16.1\% of hydroxide ions that accept 3 HBs.
The above analysis suggests that although both hydroxide ions predicted by SCAN and PBE0-TS tend to
form the stable hyper-coordinated solvation structure, the distributions of formed HBs still exhibit some deviations.
%
%
%

\subsection{Solvation Structures}


%
%
%

\begin{figure}
  \includegraphics[width=8.5cm]{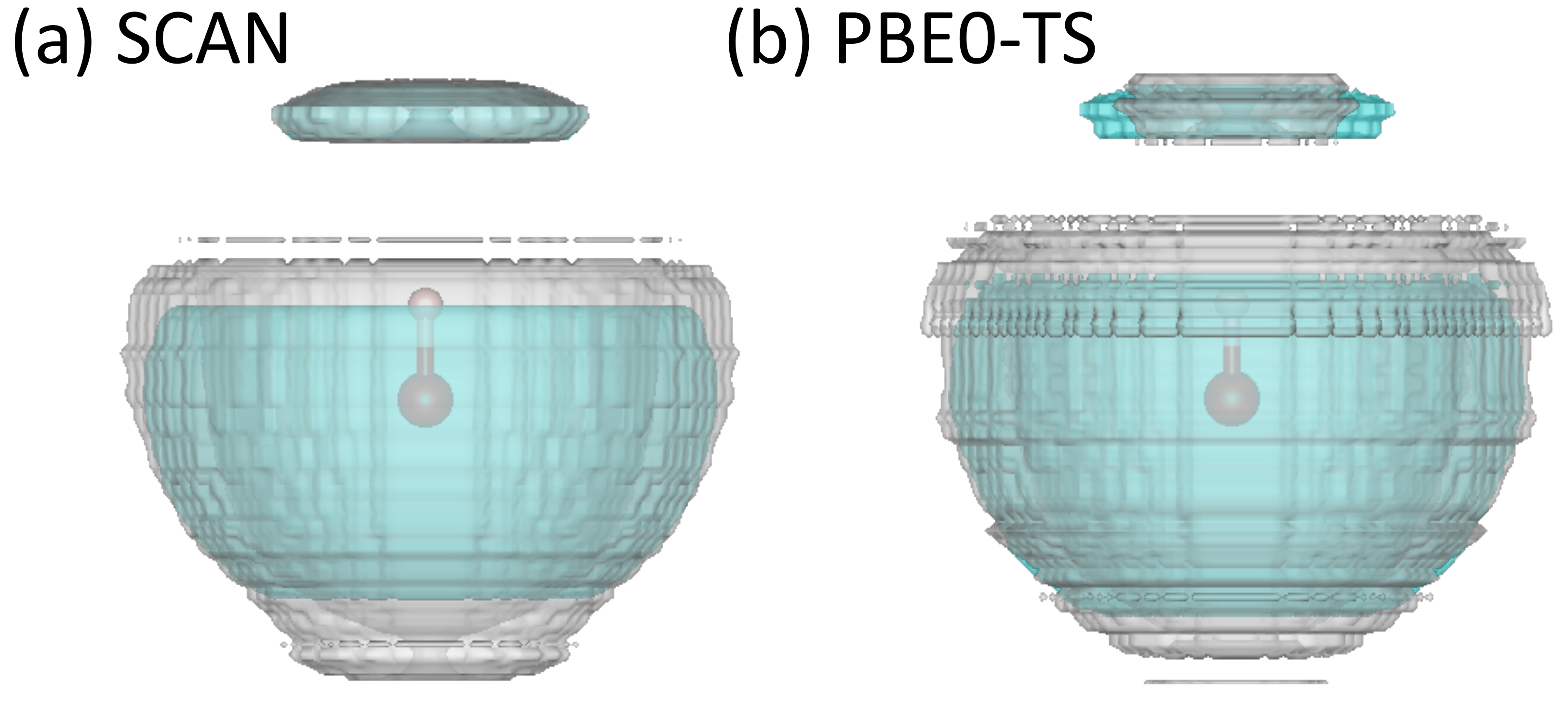}
  \caption{
 (Color online) Solvation structures of $\rm OH^{-}$(aq) in ambient liquid water as computed from AIMD trajectories utilizing the (a) SCAN and (b) PBE0-TS functionals, respectively.
  The cyan and grey isosurfaces represent the \MC{probability density plots of the distribution of oxygen atoms which are H-bonded to the} $\rm OH^{-}$(aq) ion when it accepts 4 and 5 HBs, respectively. The red and white atoms depict the oxygen and hydrogen atoms, respectively.
  }  \label{fig3}
\end{figure}

Early AIMD simulations predicted the solvation structure of the hydroxide to be planar-like.\cite{06ACR-Tuckerman}
However, a pot-like solvation structure of hydroxide was suggested by the neutron diffraction experiment coupled with empirical potential structure refinement.~\cite{03JCP-Botti}
\MC{Fig.~\ref{fig3} illustrates the probability density plot of the spatial distributions of the first solvation shell oxygen atoms that are H-bonded to the hydroxide ion. The results obtained from the SCAN and PBE0-TS functionals are shown in Figs.~\ref{fig3}(a) and (b), respectively.}
From the figure, we can see that the usage of both functionals leads to
a pot-like shape of the solvation structure, which is close to the experimental data.~\cite{03JCP-Botti}

In comparison with the pot-like shape of the solvation structure from PBE0-TS,
SCAN also provides a pot-like shape of the solvation structure but with a thinner bottom,
which is closely related to the number of accepted H-bonds.
As previously discussed, SCAN yields a larger portion of hydroxide ions which accept 5 HBs as compared to PBE0-TS.
In this regard, we examine the solvation structure features of hydroxide with different coordination numbers as described by the two methods,
in order to clarify
how different XC functionals affect the solvation structure in detail.
Figs.~\ref{fig3}(a) and (b) plot the solvation structures of hydroxide
in terms of accepting 4 (cyan) or 5 (grey) HBs in liquid water.
For both functionals, we find that the pot-like solvation structure with 5 HBs
occupies a larger volume as compared to the 4 HBs counterparts,
especially on the top and bottom areas of the pot-like solvation structure.
This indicates that the accepted fifth oxygen atom locates further away
from the hydroxide ion.
Interestingly, when accepting its fifth H-bond,
the hydroxide described by SCAN tends to locate the HB at the bottom of the `pot',
while PBE0-TS tends to place the HB on the top.
It is worth mentioning that while the excessive HB does not influence the distribution of donated HB
in the SCAN trajectory, it substantially affects the distribution of donated HB in the PBE0-TS trajectory,
as Fig.~\ref{fig3}(b) shows that the lid of the pot becomes thinner when the hydroxide accepts 5 HBs as compared to 4 HBs.

\begin{figure}
  \includegraphics[width=8.5cm]{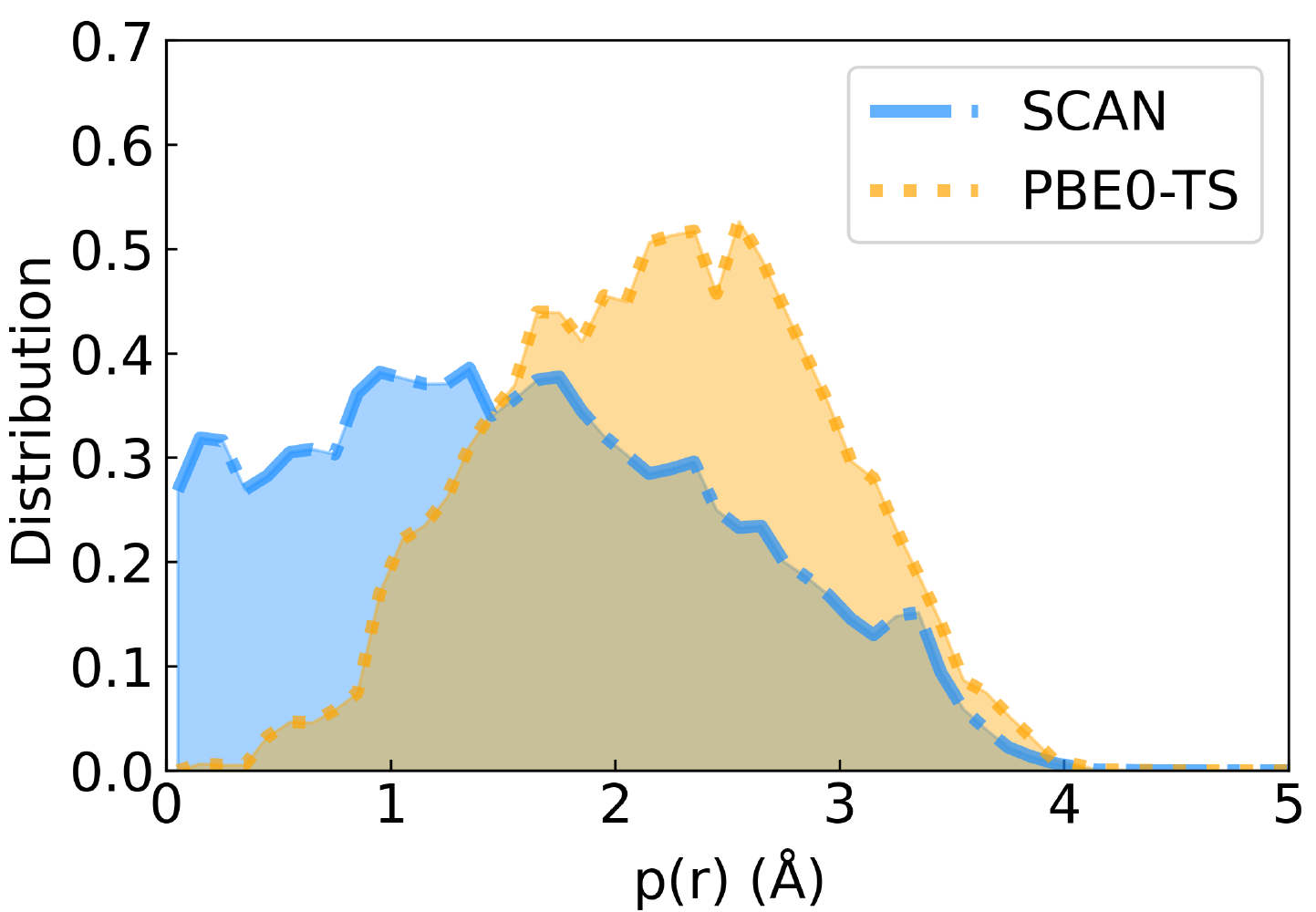}
  \caption{
  (Color online) Distributions of the planarity order parameter calculated with SCAN and PBE0-TS trajectories
  are shown.
  The planarity order parameter $\mathrm{p(r)}$ is defined as the distance from an oxygen atom
  to the plane formed by the other three oxygen atoms in water molecules.
  }  \label{fig4}
\end{figure}

We further compare the planarity order parameter $\rm p(r)$ to understand the local solvation structure of hydroxide that accepts 4 HBs. The parameter $\rm p(r)$ is defined as the distance from the fourth accepted oxygen atom of hydroxide to the plane formed by the other three oxygen atoms.~\cite{18NC-Chen} Fig.~\ref{fig4} illustrates the $\rm p(r)$ of hydroxide from the SCAN and PBE0-TS trajectories. The two trajectories lead to substantial differences in $\rm p(r)$.
For example, $\rm p(r)$ of SCAN exhibits a notable distribution from zero to around 2.0~\AA,
while $\rm p(r)$ of PBE0-TS owns a peak located at a larger planarity.
Essentially, the results suggest that the hydroxide ion described by SCAN gives rise to a more planar distribution of accepted O atoms,
while the PBE0-TS softens the directional HB strength by mitigating self-interaction error and forms a pot-like solvation structure.
The analysis is in accordance with the above comparison made between the cyan isosurfaces in Figs.~\ref{fig3}(a) and (b).
Notably, the pot-like solvation structure by SCAN relies more on the relatively abundant 5-coordinated HB structure,
which yields a solvation structure with a thinner bottom as compared to PBE0-TS.
In regard to the relationship between solvation structure and diffusion, a more planar geometry of the solvation structure of hydroxide deviates more from a tetrahedral HB network of water than PBE0-TS, resulting in a more stable asleep mode of hydroxide.

\subsection{Electronic Structures}

\begin{figure}
  \includegraphics[width=8.5cm]{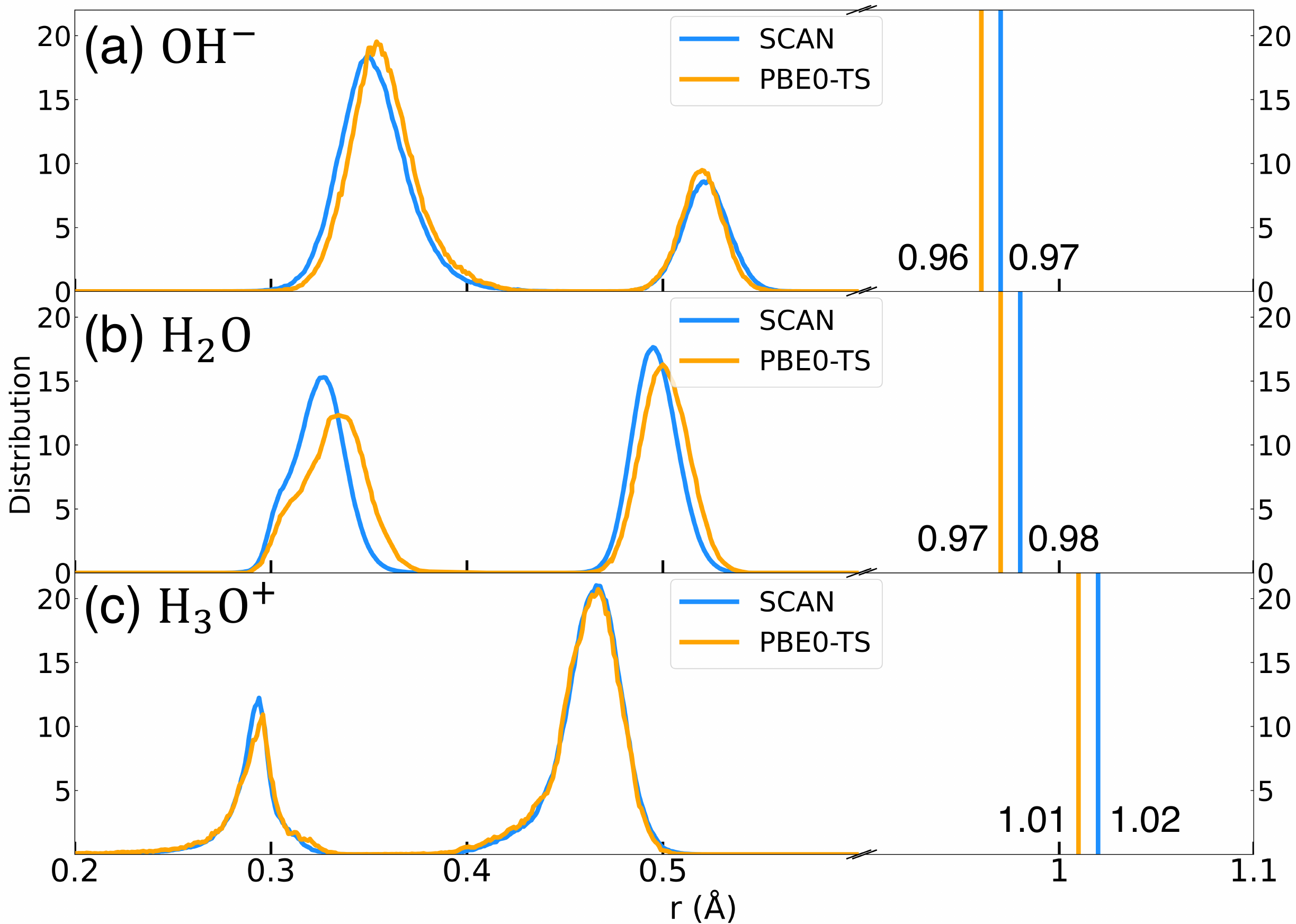}
  \caption{
  (Color online) (a), (b), and (c) respectively illustrate the distributions between the oxygen atoms of $\rm OH^{-}$(aq),
  $\rm H_{2}O$, and $\rm H_{3}O^{+}$(aq) molecules and the centers of the corresponding
  maximally localized Wannier functions.
  Blue and orange lines represent the data extracted from the SCAN and PBE0-TS trajectories, respectively.
  Vertical lines represent the average lengths of the O-H covalent bond.
    }  \label{MLWF}
\end{figure}

\begin{figure}
  \includegraphics[width=8.5cm]{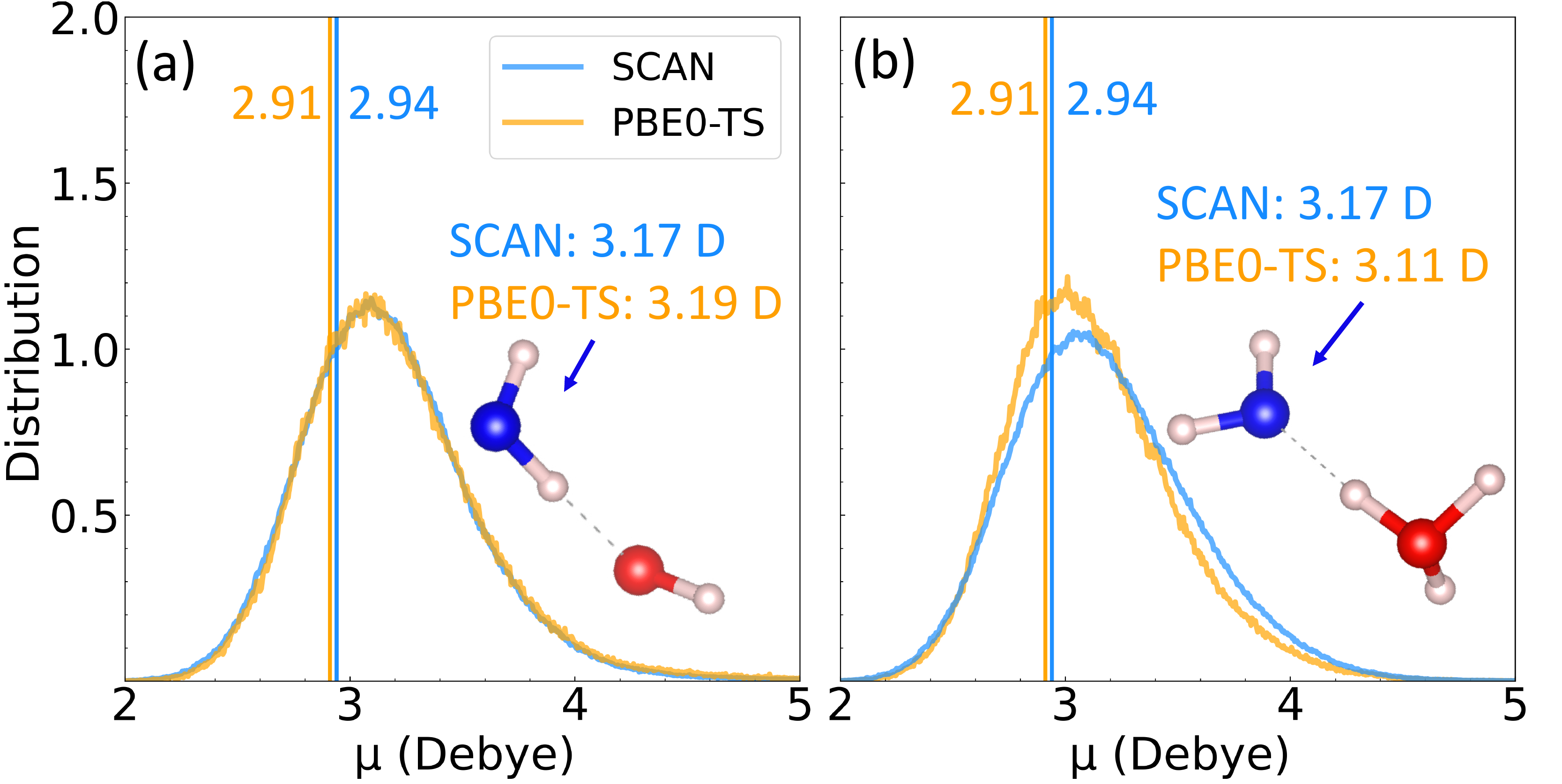}
  \caption{ 
  (Color online) \MC{Distributions of dipole moments magnitudes ($\mathrm{\mu}$) of water molecules (in blue) H-bonded to the hydrophilic end of ions (in red) are shown for (a) hydroxide and (b) hydronium ions.
  Blue and orange curves represent distributions from the SCAN and PBE0-TS trajectories, respectively.
  The average dipole moments from the two exchange-correlation functionals are shown for both ions. Vertical lines in blue and orange colors represent the average dipole moments of bulk water from the SCAN (2.94 D)~\cite{18JCP-Zheng} and PBE0-TS (2.91 D)~\cite{14JCP-Distasio} functionals, respectively.
  }
  }  \label{dipole}
\end{figure}

The hydrogen bond is largely comprised of electrostatic attractions,
which is determined by the electronic structures from DFT calculations.
In AIMD, the local electronic structure associated with chemical bonds can be conveniently described by the maximally localized Wannier functions (MLWFs)~\cite{97B-Marzari, 12RMP-Marzari}, which are obtained by unitary transformation of the DFT eigenfunctions. To be specific, in a molecular dynamics trajectory,
the information for the electronic environment for water molecules
can be deduced from the statistical analysis of the MLWF centers.
Figs.~\ref{MLWF}(a), (b), and (c) show the distributions of centers of
MLWFs representing electronic information in
$\rm OH^{-}$(aq), $\rm H_{2}O$, and $\rm H_{3}O^{+}$(aq) molecules in ambient liquid water, respectively.
The four Wannier function centers surrounding a water molecule or water ion
are divided into two peaks according to their distances \MC{from} the central oxygen atom.
The peak located closer to the O atom represents the lone pairs of electrons
while the peak further from the O atom represents the bonding pairs of electrons constituting covalent bonds.
\MC{The centers of MLWFs can be utilized to compute the molecular dipole moments with the following formula~\cite{05L-Sharma, 14JCP-Distasio, 18JCP-Zheng}:
\begin{equation}
    \mathbf{\mu} = \mathbf{R_{H_1}}+\mathbf{R_{H_2}}+6\mathbf{R_{O}}-2\sum_{i=1}^{4}\mathbf{R_{W_i}},
\end{equation}
where $\mathbf{R_{H_1}}$, $\mathbf{R_{H_2}}$ and $\mathbf{R_{O}}$ are the coordinates of the three atoms in a water molecule,
and $\mathbf{R_{W_i}}$ with $i = 1, 2, 3, 4$ are coordinates of the four MLWF centers.
In this work, we calculate the distributions of molecular dipole moment magnitudes of water molecules
involved as donor (acceptor) on the hydrophilic end of $\mathrm{OH^{-}}$ ($\mathrm{H_{3}O^{+}}$). 
Fig.~\ref{dipole} illustrates the results from both SCAN and PBE0-TS functionals, as well as the dipole moments for pure liquid water obtained from other works.~\cite{14JCP-Distasio, 18JCP-Zheng}
Note that for the hydrophobic sites of the two ions, 
we observe that the neighboring water molecules are less affected by the presence of water ions and their dipole moments are close to those in bulk water.
}
%
%

%
%

The positions of the lone pairs and bonding pairs of electrons determine the amphiphilic nature of molecules.
Taking the SCAN trajectories as an example, Figs.~\ref{MLWF}(a) and (b) illustrate that
the length of covalent bond in hydroxide and ambient liquid water, i.e., the distance between the oxygen and the proton, is 0.97 and 0.98 $\rm \AA$, respectively.
Meanwhile, the distance between the oxygen atom and the covalent bonding pairs
in hydroxide and in liquid water is 0.52 and 0.49 $\rm \AA$, respectively.
Therefore, the distance between the proton and the electron bonding pairs is smaller in hydroxide (0.45 $\rm \AA$) than in liquid water (0.49 $\rm \AA$),
leading to a less positively charged environment of hydrogen in hydroxide. The above explains the origin of the hydrophobic site of the hydroxide anion, where the formation of HBs is hindered due to the weak electrostatic attraction force.
On the other hand, the lone pair of hydroxide locates further away from the oxygen atom (0.35 $\rm \AA$) than the lone pairs of liquid water molecules (0.33 $\rm \AA$),
which brings the hydroxide ion a more negatively charged local environment around its oxygen site. Consequently, this leads to the hydrophilic property of hydroxide anion, where acceptance of HBs becomes easier than water molecules.
\MC{As illustrated in Fig.~\ref{dipole}(a),
in correspondence to the more negatively charged oxygen site, the water molecules accepted by hydroxide become more polarized with an average dipole moment of 3.17 D,
which is substantially larger than the average dipole moment of 2.94 D in bulk water.
Meanwhile, the average dipole moment of the H-bonded water on the hydrophobic end is only 2.93 D,
which is close to the average value of 2.94 D from pure liquid water.
}
In stark contrast, the electron bonding pairs of hydronium cation stay further away from the protons,
creating a more positively charged (hydrophilic) environment to stably form three HBs, as illustrated in Fig.~\ref{MLWF}(c);
the oxygen site of hydronium stays closer to the lone pair, leading to a less negatively charged environment that exhibits hydrophobic properties.
\MC{
Induced by the strong positively charged hydrogen site,
the dipole moment of water molecules accepting HBs from hydronium also increases significantly.
As shown in Fig.~\ref{dipole}(b), the average dipole moment of water molecules accepting hydrogen bonds from hydronium increses from 2.94 D in bulk water to the value of 3.17 D. Meanwhile, the average dipole moment of the water molecule that accepts a hydrogen bond from the oxygen site of hydronium reaches 2.96 D, similar to the value in bulk water.
}

Now we compare the detailed electronic structure differences of the two ions from SCAN and PBE0-TS.
First, as shown in Figs.~\ref{MLWF}(a), (b), and (c),
a larger distance between oxygen and proton (the length of O-H covalent bond)
suggests that oxygen binds more loosely with the proton,
and SCAN is known to strengthen the covalent bonds as compared to PBE~\cite{17PNAS-Chen};
here we find PBE0-TS yields protons that locate closer to the oxygen atom,
implying a stronger covalent bond as compared to SCAN.
Besides, for the pure water described by the SCAN functional,
both lone pairs and bonding pairs of ambient liquid water molecules shown in Fig.~\ref{MLWF}(b)
shift closer to the oxygen atom.
This indicates that SCAN provides a less negatively charged environment
on the oxygen site and a more positively charged environment on the proton side for pure liquid water.
In other words, the oxygen and proton sites of water molecules respectively
become more hydrophobic and more hydrophilic in the SCAN trajectory than in the PBE0-TS counterpart.
The two competing effects lead to a more pronounced first peak of radial distribution functions as observed in Figs.~\ref{pdf}(c) and (d), which implies that SCAN provides a more strengthened HB at a short-range scale when compared to PBE0-TS.
\MC{In addition, when comparing the electronic structure of pure liquid water, the distribution of MLWF centers shown in Fig.~\ref{MLWF}(b) becomes more delocalized for the PBE0-TS trajectory than the SCAN trajectory, which is consistent with the effect of mitigating the self-interaction errors as provided by PBE0-TS. However, the differences of MLWF distribution for the two ions, as inferred from Figs.~\ref{MLWF}(a) and (c) by the two functionals, are not substantial.}

%
Second, for the hydroxide anion illustrated in Fig.~\ref{MLWF}(a),
the location of the electron bonding pair is slightly further away from the central oxygen,
as predicted by SCAN and compared to that from PBE0-TS.
Nevertheless, the proton locates further away from the bonding pair as described by SCAN,
suggesting a stronger hydrophilicity of the originally hydrophobic proton site in hydroxide.
This leads to an increased number of donated HBs from PBE0-TS (0.471) to SCAN (0.577), as abovementioned.
For the other site of the hydroxide anion (oxygen),
when comparing the results from SCAN to those from PBE0-TS,
the former provides a shortened distance between the lone pair and the oxygen atom,
creating a more hydrophobic oxygen site of hydroxide.
However, a competing effect arises due to the usage of SCAN, i.e., the protons of water molecules
become more hydrophilic. As a result, the oxygen site of hydroxide in the SCAN trajectory accepts more HBs (4.101)
than that of PBE0-TS (3.922), as listed in Fig.~\ref{amph}.
The above differences that exist in the electronic structures of hydroxide anion
lead to the deviation of the solvation shell structure \MC{obtained from the SCAN and PBE0-TS trajectories, as illustrated in Fig.~\ref{fig3}.}
\MC{As for the average dipole moment of water molecules that donate H-bonds to the hydroxide ion,
Fig.~\ref{dipole}(a) shows that the values are 3.17 and 3.19 D from the SCAN and PBE0-TS exchange-correlation functionals, respectively.
We conclude that the two functionals yield similar dipole moments for the water molecules at the hydrophilic end of the hydroxide ion.}
\RX{
%
%
%
%
%
%
}
%
%

Third, by analyzing the distribution of MLWF centers of hydronium cation from the SCAN and PBE0-TS trajectories shown in Fig.~\ref{MLWF}(c), we observe that SCAN provides a more hydrophilic environment for the proton sites of hydronium because the distance between the bonding pair and the protons is larger. In addition, the protons of hydronium interact with the more hydrophobic oxygen of water molecules. As a result, hydronium from both trajectories donates three stable HBs, suggesting that SCAN and PBE0-TS yield similar short-ranged structure information for the proton sites of hydronium.
\MC{In addition, Fig.~\ref{dipole}(b) shows that the two functionals predict similar dipole moments for the water molecules that accept H-bonds from the hydronium ion; Specifically, the value is 3.17 and 3.11 D from SCAN and PBE0-TS, respectively.}
%
%
%
On the other hand, for the oxygen site of hydronium, the lone pair from SCAN locates slightly closer to the oxygen atom as compared to the one from PBE0-TS, which creates a slightly more hydrophobic environment for hydronium. In this regard, it is consistent that hydronium cation accepts 0.138 HBs from SCAN, which is marginally smaller than the 0.140 HBs from PBE0-TS, as listed in Fig.~\ref{amph}.

In conclusion, for water ions including the hydronium and hydroxide ions,
as well as the water molecules, we analyze their electronic structures via the distribution of MLWF centers.
In general, compared to the results from PBE0-TS,
we find that the SCAN functional yields a closer distance between the lone pairs and the oxygen atom,
and a larger distance between the bonding pairs and protons.
In addition, the hydrophilic and hydrophobic nature of the two ions are affected by considering the delicate interactions between the ions and neighboring water molecules. Consequently, when comparing the SCAN results with the PBE0-TS results, the hydroxide anion from SCAN becomes more hydrophilic, while the hydronium cation from SCAN becomes slightly more hydrophobic.

\subsection{Proton Transfer}

\begin{figure*}
  \includegraphics[width=17cm]{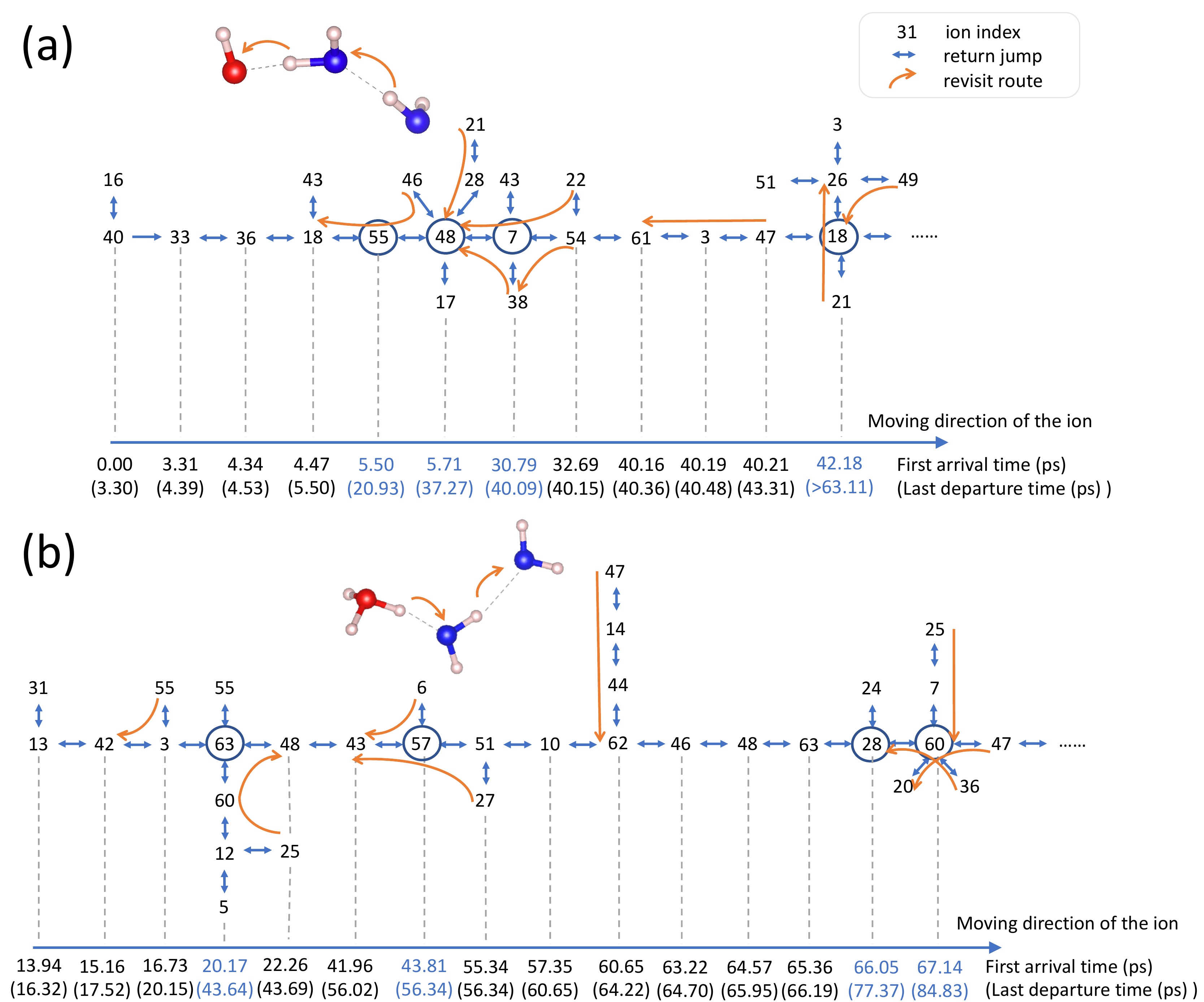}
  \caption{\RX{
  \RX{(Color online)} Index of ionized water molecules with respect to time in the SCAN AIMD trajectory for (a) hydroxide and (b) hydronium ions. The blue double-headed arrow stands for the forward and return jumps between two adjacent water molecules. The orange arrow stands for the revisit route. The ion indices with blue circles stand for the central water molecules where proton (hole) is trapped. The time of first arrival and last departure of the proton (hole) is shown below the time axis.}
    }  \label{ion_index}
\end{figure*}

\begin{figure}
  \includegraphics[width=8.5cm]{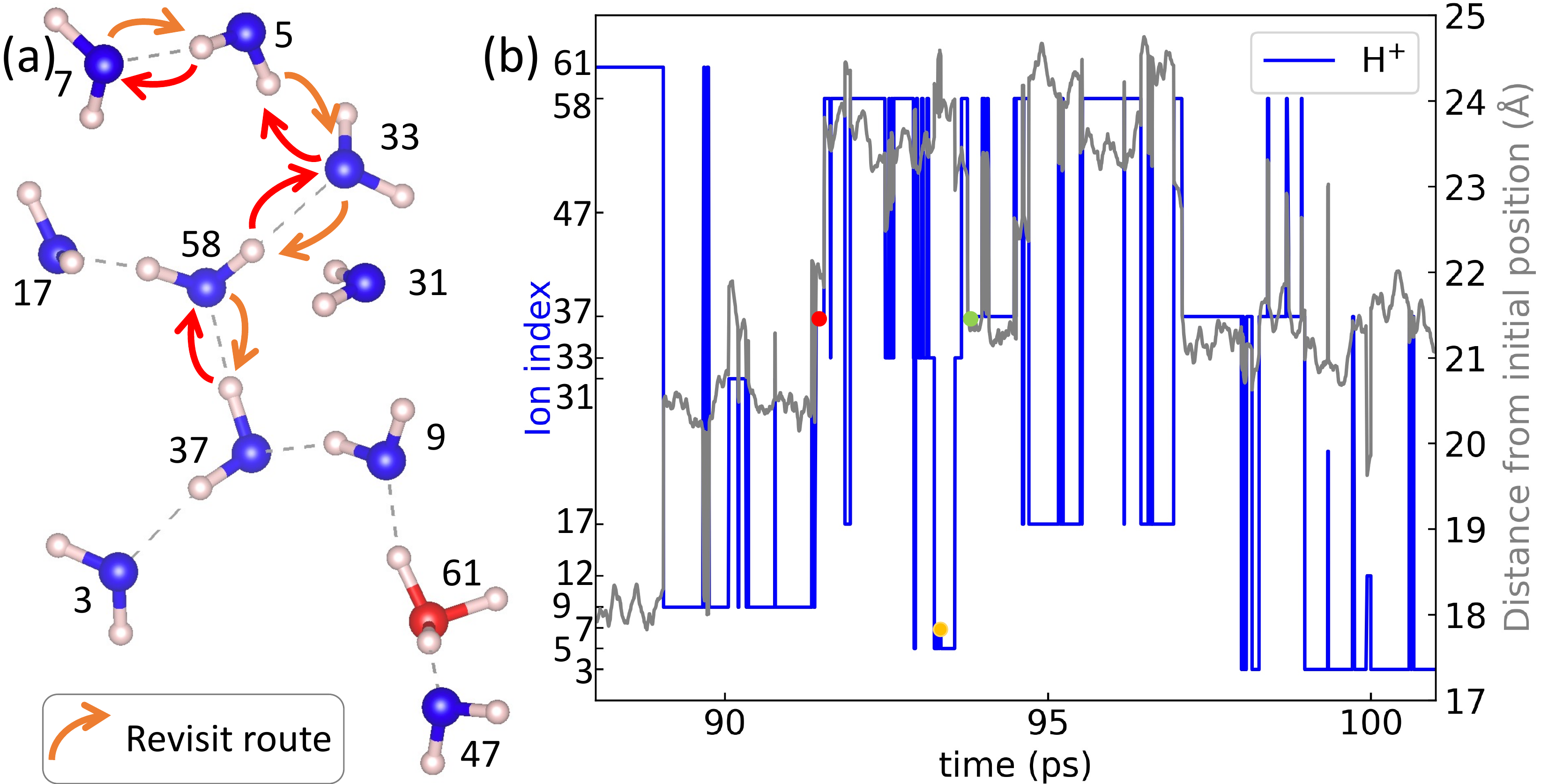}
  \caption{
  \RX{(Color online)} \RX{(a) Schematic plot of the revisit route for hydronium consists of five water molecules with ion indices being 37, 58, 33, 5 and 7. 
  (b) Changes of ion index (blue) and distance from initial position (grey) with respect to time. The first and second time that the excess proton arrives at water with index of 37 is marked with red and green points, respectively, while the time that the ion arrives at water with index of 7 is marked with an orange point. }
    }  \label{long_return}
\end{figure}

\RX{Recently, the PT mechanism of the two ions has received increasing attention from simulations.
For example, the diffusion coefficients of the two ions are largely determined by the PT phenomenon. In fact, only part of the PT events contributes to the final diffusion coefficients, the rest are usually classified as non-diffusional PTs such as rattling.~\cite{18NC-Chen, 07L-Chandra, 18CS-Fischer, 15JCP-Tse, 21JCP-Arntsen}
However, defining the rattling events are non-trivial.~\cite{18CS-Fischer,21JCP-Arntsen}
Until now, several classification methods of PT events exist due to the complex behavior of the PT phenomenon itself.~\cite{07L-Chandra, 15JCP-Tse, 18CS-Fischer, 13PNAS-Hassanali} For instance, Hassanali et al.~\cite{13PNAS-Hassanali} proposed to divide the PT events into single and concerted jumps, the latter of which dominate the PT phenomenon for both hydronium and hydroxide ions. Later, the concerted jump mechanism was revised to be dominant only in hydronium by using a more accurate XC functional PBE0-TS in AIMD simulations.~\cite{18NC-Chen} Recently, Arntsen et al. suggested that the concerted jumps of hydronium were in fact dominated by non-diffusional jumps.~\cite{21JCP-Arntsen}
In addition, it is worth mentioning that the `special pair dance' (SPD) concept, which was proposed for hydronium,~\cite{08JPCB-Omer} refers to the period during which no PT occurs and
the closest H-bonded neighbor of the protonated water molecule quickly interchanges
at a frequency of tens of femtoseconds. In this regard, the SPD concept has no overlap with the classification of PT.}
%
\RX{
As a typical phenomenon of the non-diffusional PTs, water ions tend to be trapped among a few water molecules and keep rattling between them, as explained by previous studies using either ring distribution\cite{13PNAS-Hassanali} or accepted HB of hydronium.~\cite{15JCP-Tse}
We also observe similar trapping phenomena in the SCAN trajectories.
Figs.~\ref{ion_index}(a) and (b) respectively illustrate the representative PT events for hydroxide and hydronium from the SCAN trajectories, where the change of ion indices with respect to simulation time is plotted.
The trapping events are identified as blue circles in Fig.~\ref{ion_index}, from which the proton (hole) tends to hop among neighboring water molecules with a high frequency for more than a few ps but contribute little to the diffusion coefficients.}

\RX{
In order to remove the non-diffusional PTs, Arntsen et al.~\cite{21JCP-Arntsen} proposed to apply a filter to the molecular dynamics trajectories. The filter eliminates the short-ranged hopping events that are followed directly by a back hop returning to the original water. In other words, two molecules were used to define the return events. The method can also be applied to trim the ``slingshot" effect which involves more than two water molecules but cannot fully eliminate the non-diffusional PTs.
Fischer et al.~\cite{18CS-Fischer, 19JPCB-Fischer} 
inherently defined the PT events within the first solvation shell of hydronium as rattling, which typically involves three or four water molecules within the Eigen complex $\mathrm{H_{3}O^{+}[H_2O]_3}$.
We find similar rattling effects in the SCAN trajectory. Interestingly, we observe additional long-ranged return events involving more than four water molecules, which do not have to rely on the concept of Eigen complex.
Specifically, the long-ranged return phenomenon refers to a water ion hops for several steps and occasionally revisits its former hosts by following exactly the same route. %
In the following, we refer to these long-ranged return events that involve three or more water molecules as the `revisit' events.}
\RX{
Fig.~\ref{ion_index} illustrates multiple revisit routes (orange arrows) as obtained from the SCAN trajectories, which suggest that non-diffusional events are in fact a complex phenomenon that involves several water molecules.
As illustrated in Fig.~\ref{ion_index}(b), several revisit routes of hydronium involve three water molecules, most of which result from frequent switches of proton rattling partners.
For example, in the first revisit event of hydronium, the proton first rattles between water molecules with indices being 3 and 42, then between water 3 and 55, followed by rattling again between water 3 and 42.
Another long-ranged revisit event is shown in Fig.~\ref{long_return}, where the revisit route
is shown in Fig.~\ref{long_return}(a) while the changes of ion index and distance from the initial position with respect to time are plotted in Fig.~\ref{long_return}(b).
We can see that
after the proton hop from the water molecule with index of 9 to another one with the index of 37, the proton takes a long journey by visiting five different water molecules with indices being 37, 58, 33, 5 and 7.
The time of starting and ending points is respectively marked as red and orange points in Fig.~\ref{long_return}(b).
After arriving at water with index of 7,
the excess proton rapidly goes back to water with index of 37 by following the same route ($7\rightarrow5\rightarrow33\rightarrow58\rightarrow37$). The return point is marked as a green point in Fig.~\ref{long_return}(b).
By observing the change of distance in Fig.~\ref{long_return}(b) after the revisit event ends, we find these PT events contribute little to the diffusivity of hydronium.
%
%
In summary, the existence of the above two non-diffusional PT phenomena, i.e., trapping and revisit, demonstrates that the classification of PT events needs more investigations in future.
}

\begin{figure}
  \includegraphics[width=8.5cm]{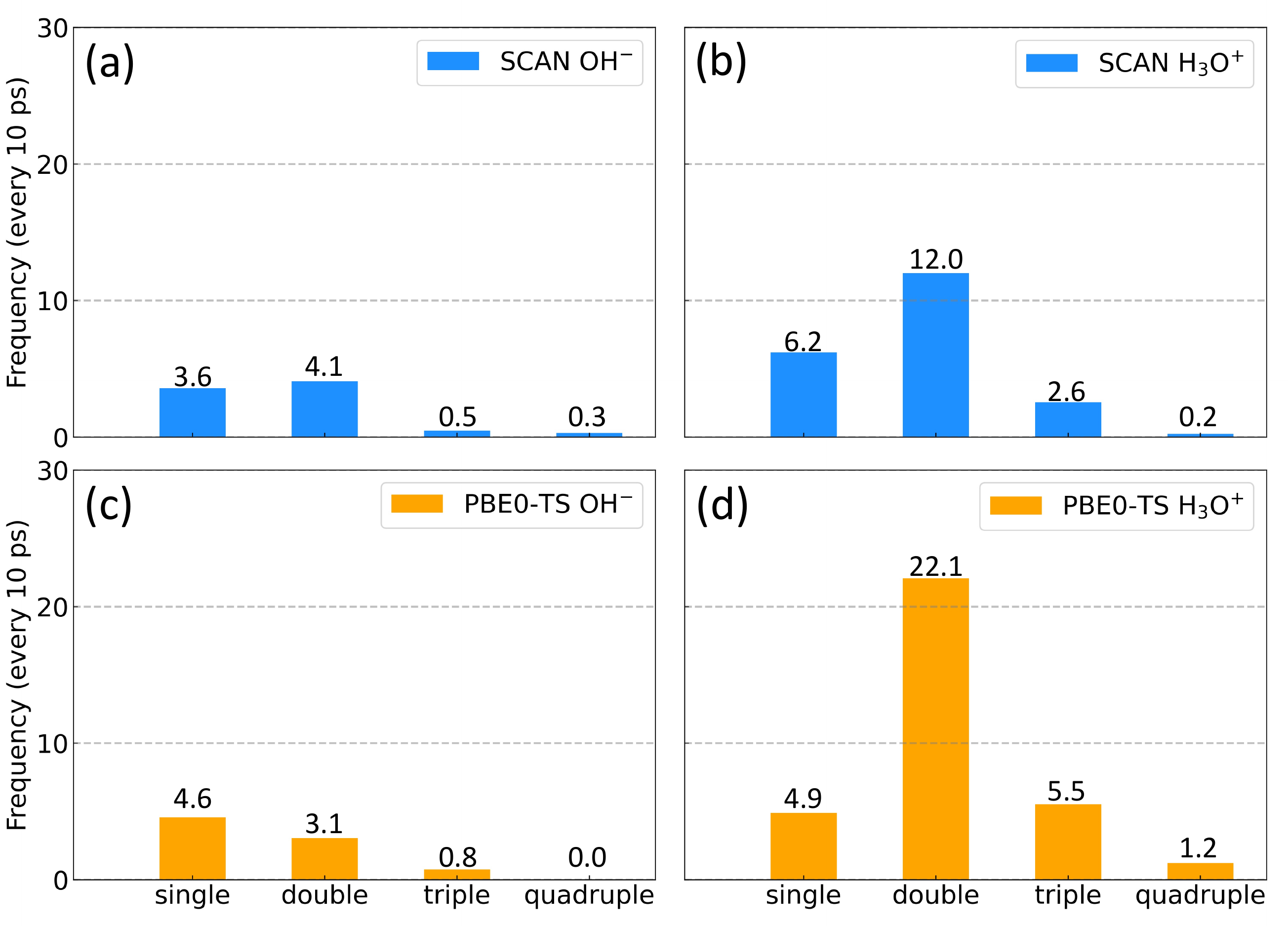}
  \caption{
  (Color online) \RX{Proton transfer events are classified into four different types, i.e., single, double, triple, and quadruple jumps. Note that the definition of PT events includes both diffusional and non-diffusional PT events. \RXN{Results are obtained for (a) $\rm OH^{-}$(aq), (b) $\rm H_{3}O^{+}$(aq) from the SCAN functional and (c) $\rm OH^{-}$(aq), (d)  $\rm H_{3}O^{+}$(aq) from the PBE0-TS functional.}}}  \label{MJ}
\end{figure}

\begin{figure}
    \centering
    \includegraphics[width=8.5cm]{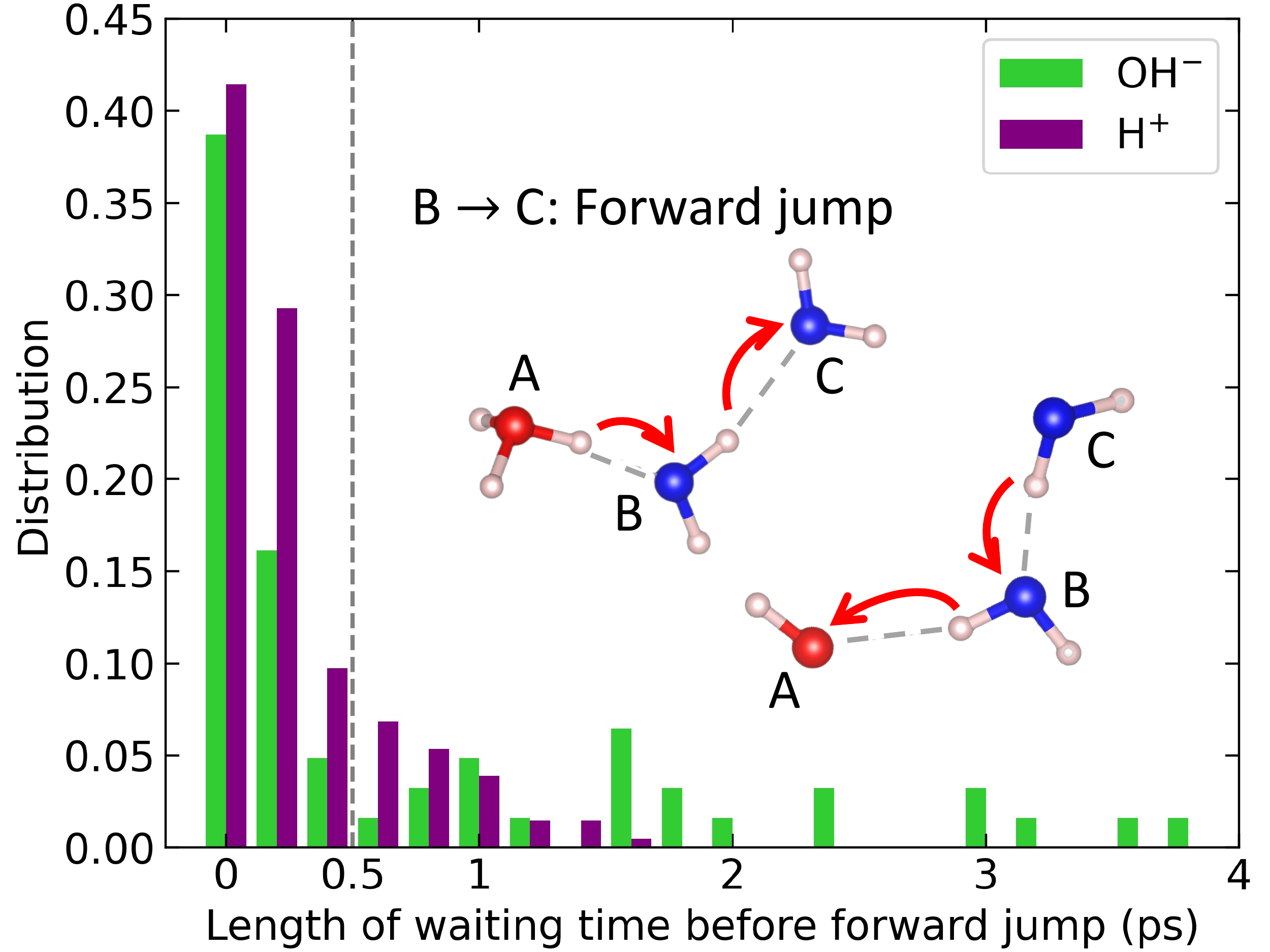}
    \caption{\RX{
    (Color online) Distribution of Waiting time between two successive jumps (inset figure) for proton hole (green) and proton (purple). Here the second jump has to be a forward jump to satisfy the definition of concerted jump. A 0.5 ps cutoff for the waiting time is shown in grey dashed line.}
    }
    \label{time_interval}
\end{figure}

\begin{figure}
  \includegraphics[width=8.5cm]{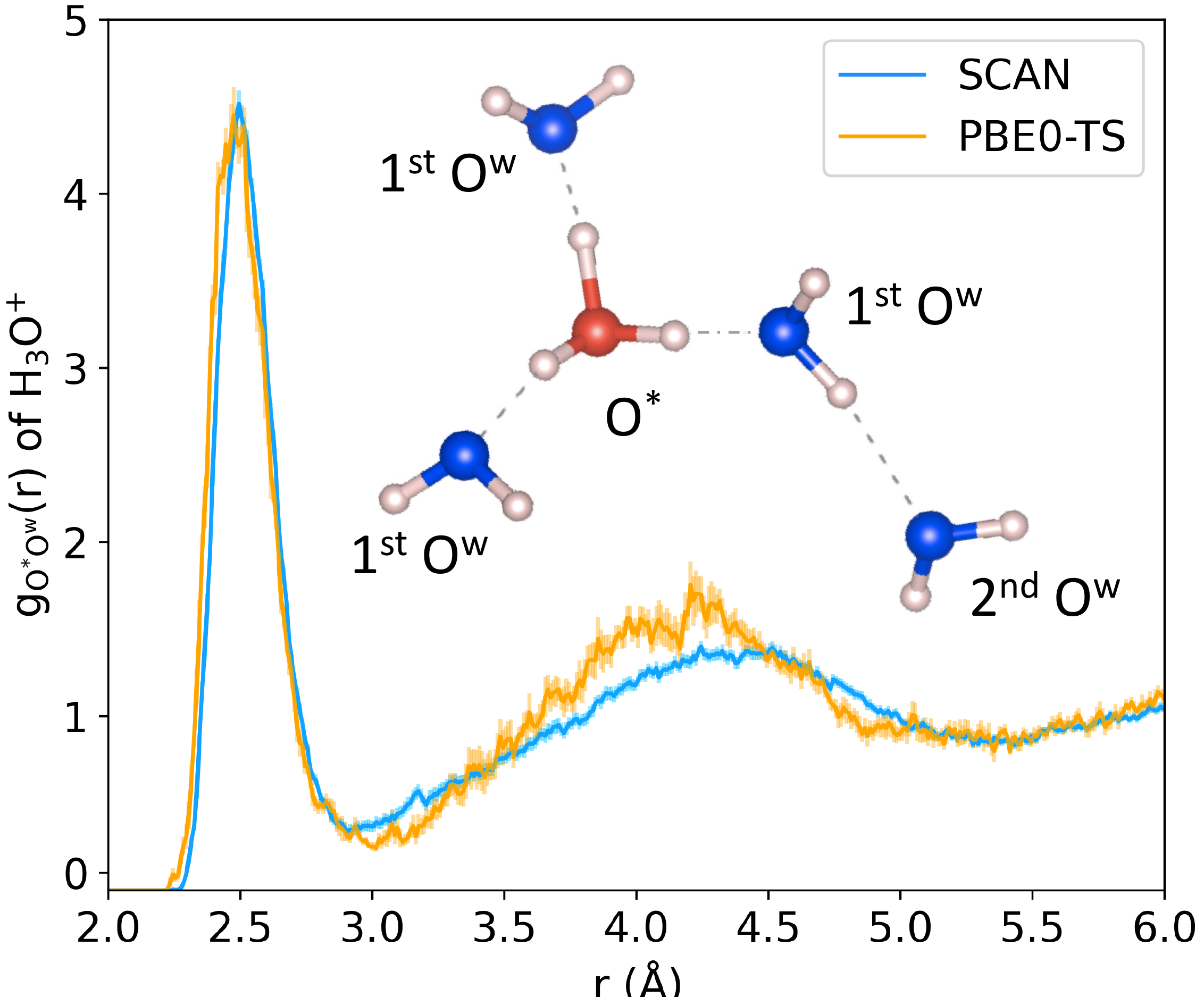}
  \caption{
 (Color online) Radial distribution functions involving O atoms in $\rm H_{3}O^{+}$ ($\mathrm{O}^{\ast}$) and O atoms from neighboring water molecules ($\mathrm{O}^{W}$). Two trajectories involving SCAN and PBE0-TS exchange-correlation functionals are used. \MC{Inset illustrates the geometrical environment of a hydronium ion ($\mathrm{O^{*}}$) and its first-shell ($\mathrm{1^{st}}$ $\mathrm{O^{w}}$) and second-shell ($\mathrm{2^{nd}}$ $\mathrm{O^{w}}$) neighboring water molecules. Error bars are illustrated for each curve in light colors.} }  \label{OxO}
\end{figure}


\RX{
Importantly, studying both non-diffusional and diffusional PTs from simulation is informative based on the following three considerations.
First, non-diffusional PT events are relevant to experimental findings. For instance, the Eigen-Zundel configuration transition of hydronium is explained by 
fast rattling between two water molecules,~\cite{68ZFPC-Zundel, 58PRSL-Eigen, 95CPL-Agmon} and recent infrared spectroscopy experiments provide new evidences to identify which species dominates the PT process.~\cite{18NC-Founier, 18JPCB-Carpenter}
Second, both non-diffusional PTs and the SPD phenomenon are the intrinsic properties of the two ions that need to be addressed from a scientific point of view. A recent work found different PT patterns exist in solvated hydronium and hydroxide ions, i.e., hydroxide experiences much less PT events than hydornium by considering both diffusional and non-diffusional PT events.~\cite{18NC-Chen} In this regard, it is worth investigating more microscopic origins to elucidate the different PT patterns of the two ions, which could be useful in designing relevant experiments. 
Third, Tse et al.~\cite{15JCP-Tse} mentioned that the frequency of concerted hops involving three water molecules depends on the exchange-correlation functional.
Therefore, in order to verify the accuracy of SCAN functional in describing both solvated hydronium and hydroxide ions, it is a necessity to carry out comprehensive tests on both structural and dynamic properties, including both diffusional and non-diffusional PTs.}

In this work, we study both diffusional and non-diffusional PTs for the two ions.
Essentially, a conceptually correct picture of the structural diffusion of $\rm OH^{-}$(aq) and $\rm H_3O^{+}$(aq) ions
relies on a precise description of the HB network of water,
as well as the solvation structures of water ions.
Although the issues are well addressed by the PBE0-TS functional,~\cite{18NC-Chen}
the computational cost is extremely high.
As abovementioned, the number of formed HBs between molecules stands out as a major indicator
of hydrophobicity and hydrophilicity.
Generally dictated by electronic structures,
the HB network of water and ions is mainly determined by the used XC functional.~\cite{92CPL-Laasonen, 93CPL-Laasonen, 16JCP-Gillan, 15JPCL-Gaiduk,
09JPCB-Schmidt, 11JCP-Wang, 15JCP-Miceli, 17PNAS-Chen, 14JCP-Distasio, 98-Burke, 00JCP-Boese, 06ACR-Tuckerman,19AR-Sakti}
Therefore, we validate the performance of the SCAN XC functional
in providing the frequency of the PT events 
and compare them to the PBE0-TS results.

In order to quantitatively describe the frequency of 
PT events in these AIMD trajectories,
we categorize the PT events according to the number of transfer events during one burst, 
as done in Ref.~\onlinecite{18NC-Chen}. We define a single jump as a PT event not followed by any PT events in the following 0.5 ps. The 0.5 ps is a characteristic time to compress the water wire, as introduced in Refs.~\onlinecite{11PNAS-Hassanali}.
If more than one PT event occurs within 0.5 ps and the proton does not return to any of its former host molecules in these jumps,
these PT will be labeled as a concerted jump. 
\RXN{The time cutoff of 0.5 ps is picked on an empirical basis to facilitate comparison between XC functionals,
while more physically meaningful time cutoff could be deduced from the PT correlation function.~\cite{96N-Luzar, 07L-Chandra, 09L-Berkelbach}}
\RX{
Note that the rest of PT events as labelled as simple rattling events, which are a subset of rattling events from other definitions that involve two~\cite{21JCP-Arntsen} or more molecules.~\cite{18CS-Fischer} As abovementioned, none of these definitions could fully eliminate the non-diffusional PTs. In addition, most definitions are applied to hydronium only. In this regard, further investigations for defining rattling for both hydronium and hydroxide ions are needed.}

Figs.~\ref{MJ}(a) and (c) display statistics of the concerted PT events for hydroxide by using
the SCAN and PBE0-TS functionals, respectively.
Generally, both XC functionals reveal that the total frequency of PT in hydroxide is lower than hydronium.
We also find that the two functionals yield the same conclusion that PT events in the hydroxide
system are dominated by both single and double jumps, while triple and quadruple jumps are relatively rare.
In stark contrast, as illustrated in Figs.~\ref{MJ}(b) and (d), 
where both functionals yield the same conclusion that 
the concerted PT events for hydronium is dominated by the double jumps,
the frequency of which is substantially larger than the counterpart in hydroxide.
For instance, the frequency of double jumps from SCAN (PBE0-TS) is 12.0 (22.1) times per 10 ps for hydronium, 
much larger than the value of 4.1 (3.1) for hydroxide.
In addition, the triple- and quadruple-jumps are relatively rare in hydronium.

\RX{
We define a forward jump of proton as jumping to a neighboring water molecule that is not the host in the previous jump. 
In this regard, for any PT events that concern two or more jumps, the second jump must be a forward jump before it can be classified into a multiple jump.
It is useful to collect the waiting time between the two successive jumps for the PT events of the two ions, and the results are shown in Fig.~\ref{time_interval}.
The data in Fig.~\ref{time_interval} are collected from the SCAN trajectories but similar results can be deduced from the PBE0-TS trajectories.
As Fig.~\ref{time_interval} shows, 76\% and 58\% of the forward jumps occur within 0.5 ps after the first jumps for hydronium and hydroxide, respectively.
Besides the water wire compression time of 0.5 ps as suggested in Ref.~\onlinecite{11PNAS-Hassanali}, 
the analysis here also implies that 0.5 ps is a
reasonable time scale to include most forward jumps for the two solvated ions in liquid water.
Additionally, we observe the hydroxide ion tends to wait longer before the second jump than the hydronium ion. For instance, the longest waiting time is 7.1 and 1.7 ps for the hydroxide and hydronium ions, respectively.
These results demonstrate the inactive nature of hydroxide compared to hydronium,
which agrees with the different PT frequencies of the two ions as illustrated in Fig.~\ref{MJ}.
}

\RX{
Since only the simplest rattling event is included in this classification, i.e., rattling between two water molecules, the single and multiple PTs defined here could contribute little to diffusion. We find that hydronium undergoes more simple rattling events than hydroxide, which could be attributed to the fast alternate between the 
Zundel ($\rm H_{5}O_{2}^{+}$) and Eigen ($\rm H_{9}O_{4}^{+}$) configurations.~\cite{10CR-Marx, 06CPC-Marx}
For example, we observe the hydronium ion rattles between two water molecules continuously for at most more than 30 times within a time period of approximately 5 ps; meanwhile, the hydroxide ion in the SCAN trajectory rattle for at most 8 times between two water molecules for no longer than 2 ps.
This result demonstrates a major difference in PT pattern between hydroxide and hydronium.
}

We find SCAN yields a substantially fewer number of double, triple, and quadruple PT hops than PBE0-TS, as shown in Figs.~\ref{MJ}(b) and (d). For instance,
the frequency of double jumps from SCAN (12.0) is about half of that from PBE0-TS (22.1). Consequently,
we obtain a relatively lower diffusion coefficient of hydronium from SCAN as compared to those from PBE0-TS.
The reason can be analyzed via the three water molecules (form a water wire) involved in the double jump. In fact, the third water molecule most likely comes from the second nearest neighbor of the first water molecule.

In this regard, we plot $\rm g_{O^{*}O^{w}}(r)$ 
from the two trajectories (SCAN and PBE0-TS) in Fig.~\ref{OxO}.
First, the two functionals yield similar average dipole moment (3.17 D from SCAN and 3.11 D from PBE0-TS, in Fig.~\ref{dipole}) of the first-shell neighboring water molecules of hydronium; the increased dipole moments as compared to water originate from the hydrophilic end of the hydronium ion, forming three strong H-bonds that enable the PT events.
Second,
we can see that the first peak of $\rm g_{O^{*}O^{w}}(r)$ from PBE0-TS
is closer to the central oxygen atom of hydronium as compared to SCAN.
This indicates that SCAN gives rise to a slightly more hydrophobic hydronium than PBE0-TS, 
in agreement with previous analysis. 
As a result, the hydronium described by PBE0-TS is more likely
to donate its protons to its neighboring molecules, resulting in more frequent PT events.
Third, the second peak from the SCAN trajectory in Fig.~\ref{OxO} 
is significantly lower than that of PBE0-TS, suggesting a relatively lower local water density in the second solvation shell of hydronium.
The lack of enough water molecules in the second solvation shell of hydronium
prevents the protons from jumping two or more times within a burst of PT events.

\begin{figure}
  \includegraphics[width=8.5cm]{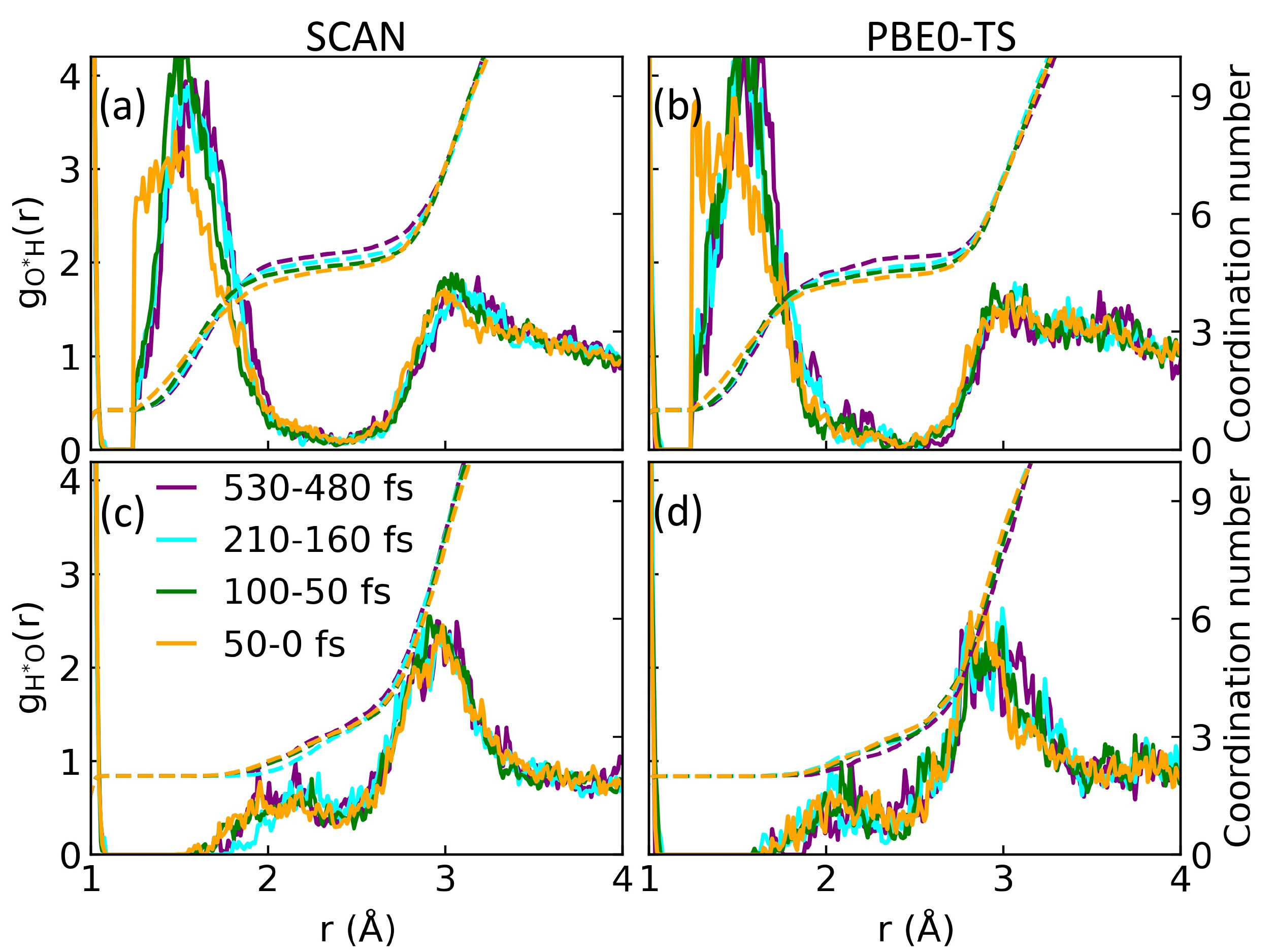}
  \caption{\RXN{
  (Color online) Radial distribution function (RDF) $\mathrm{g_{O^{*} H} (r)}$ and $\mathrm{g_{H^{*} O} (r)}$ of hydroxide in four time intervals (50 fs long for each interval) before PT generated by the SCAN and PBE0-TS functionals. RDF from 530-480 fs, 210-160 fs, 100-50 fs and 50-0 fs before PT are shown in purple, cyan, green and orange respectively. The integrated coordination numbers for the RDFs are shown in dashed lines in corresponding colors. (a) $\mathrm{g_{O^{*} H} (r)}$ and (c) $\mathrm{g_{H^{*} O} (r)}$ are obtained from the SCAN functional, while (b) $\mathrm{g_{O^{*} H} (r)}$ and (d) $\mathrm{g_{H^{*} O} (r)}$ are obtained from the PBE0-TS functional.}
    }  \label{oh_pdf_pt}
\end{figure}

\begin{figure}
  \includegraphics[width=8.5cm]{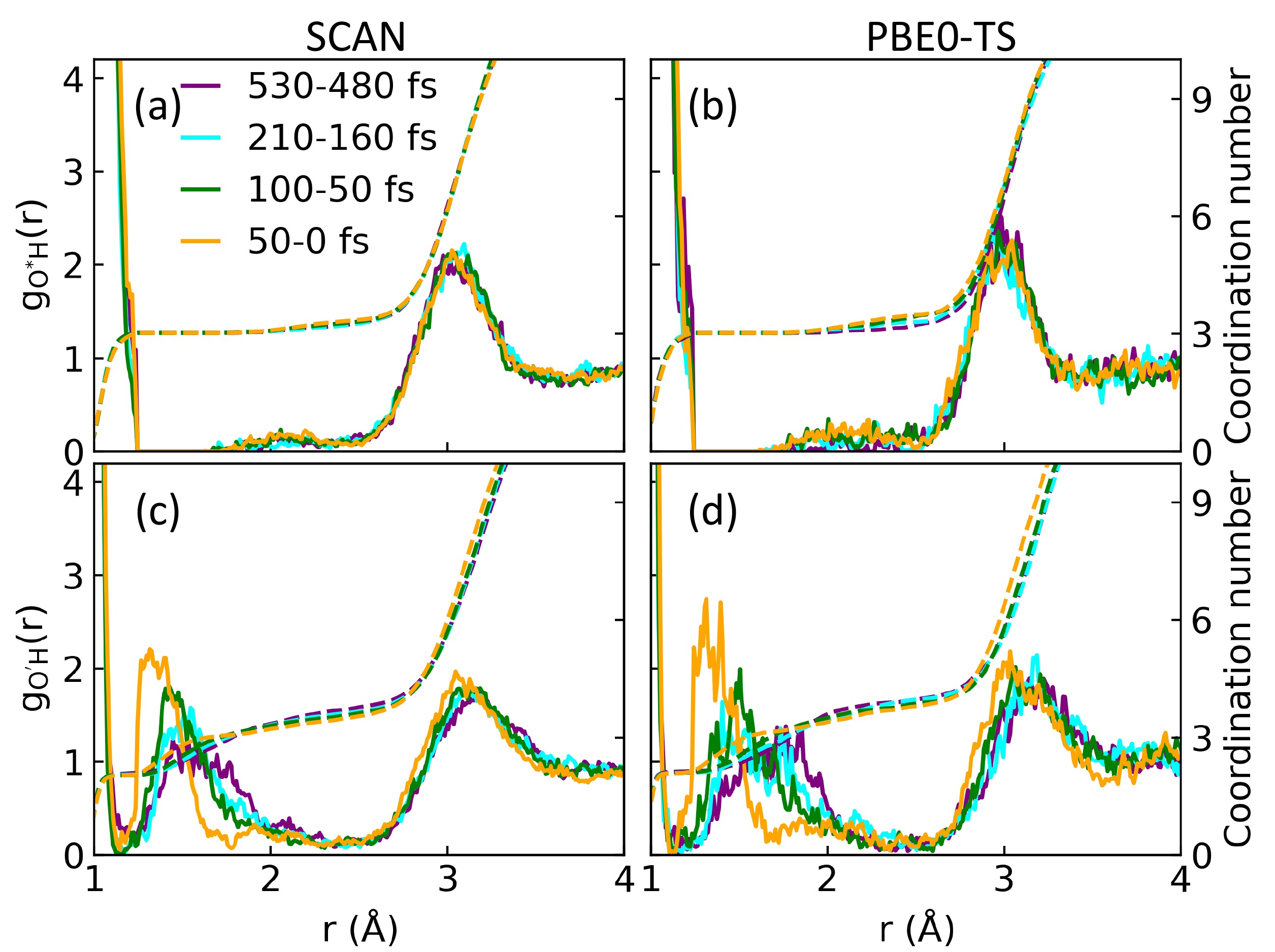}
  \caption{\RXN{
  (Color online) Radial distribution function (RDF) $\mathrm{g_{O^{*} H} (r)}$ and $\mathrm{g_{O^{'}H} (r)}$ of hydronium in four time intervals (50 fs long for each interval) before forward PT generated by SCAN and PBE0-TS functional. RDF in 530-480 fs, 210-160 fs, 100-50 fs and 50-0 fs before PT are shown in purple, cyan, green and orange respectively. Here $\mathrm{O^{'}}$ denotes the oxygen atom of the next hydronium. The integrated coordination numbers for the RDFs are shown in dashed lines in corresponding colors. (a) $\mathrm{g_{O^{*} H} (r)}$ and (c) $\mathrm{g_{O^{'}H} (r)}$ are obtained from the SCAN functional, while (b) $\mathrm{g_{O^{*} H} (r)}$ and (d) $\mathrm{g_{O^{'}H} (r)}$ are obtained from the PBE0-TS functional.}
    }  \label{h_pdf_pt}
\end{figure}

\RXN{In the end, we would like to discuss the presolvation structure with the SCAN and PBE0-TS XC functional.
Past researches have addressed s strong correlation between the PT phenomenon and the presolvation structure of ions, 
which refers to the HB structure of an ion or molecule in resemblance to the species into which it is turning after PT.~\cite{06ACR-Tuckerman}
For example, hydroxide is predicted to convert from a hyper-coordinated structure to a three-coordinated structure and donate a HB before PT.~\cite{02N-Tuckerman, 06ACR-Tuckerman, 10CR-Marx}
We follow the method of Ref.~\onlinecite{09L-Berkelbach} and plot the $\mathrm{g_{O^{*}H}(r)}$ and $\mathrm{g_{H^{*}O}(r)}$ in time periods of 530-480 fs, 210-160 fs, 100-50 fs and 50-0 fs before the PT event occurs.
On the one hand, the RDF and the integrated coordination numbers of hydroxide obtained from both SCAN and PBE0-TS trajectories are displayed in Fig.~\ref{oh_pdf_pt}.
We observe the first peak of $\mathrm{g_{O^{*}H}(r)}$ moves leftward and decreases in height when the trajectory time approaches the PT event, 
resulting in the decrease in the coordination number (Figs.~\ref{oh_pdf_pt}(a) and (b)).
For example, from the time period of 530-480 fs to 50-0 fs before PT,
SCAN (PBE0-TS) predicts the coordination number at 2.5 $\mathrm{\AA}$ decreases monotonically from 5.07 (4.90) to 4.65 (4.44),
indicating a significant loss in the accepted HB of hydroxide before PT.
The first peak of $\mathrm{g_{H^{*}O}(r)}$, however, does not render significant changes with respect to time by either functional (Figs.~\ref{oh_pdf_pt}(c) and (d)).
On the other hand, previous works have proposed that hydronium tends to accept a HB before PT. 
Meanwhile, the nearest neighbor of hydronium which is going to accept the excessive proton tends to break the other accepted HB.~\cite{09L-Berkelbach, 16JPCB-Biswas, 10CR-Marx}
In this regard, we plot $\mathrm{g_{O^{*}H}}(r)$ and $\mathrm{g_{O^{'}H}}(r)$ in time periods of 530-480 fs, 210-160 fs, 100-50 fs and 50-0 fs before the PT, which are shown in Fig.~\ref{h_pdf_pt}.
Here $\mathrm{O^{'}}$ is used to denote the oxygen atom of water molecule that accepts the excess proton.
We notice that 
the two functionals slightly diverge on the change of the coordination number of $\mathrm{g_{O^{*}H}}(r)$ from 530-480 fs to 50-0 fs before PT:
the PBE0-TS functional predicts an increasing coordination number at 2.5 $\mathrm{\AA}$ from 3.22 to 3.49,
while the SCAN functional only predicts an increase from 3.29 to 3.36.
Furthermore, 
both functionals show similar decrease of the accepted coordination number of the water molecule that accepts the excess proton:
the coordination number of $\mathrm{g_{O^{'}H}}(r)$ at 2.5 $\mathrm{\AA}$ from the SCAN (PBE0-TS) trajectory decreases from 3.77 (3.96) to 3.54 (3.79).
In summary, we find the presolvation picture is mostly valid for both functionals, except for the donation of HB by hydroxide before PT.
%
%
%
%
By using the SCAN and PBE0-TS functionals, 
we observe the occurrence of PT between a hyper-coordinated hydroxide and its neighboring water molecule or between a hydronium without accepted HB and a water with two accepted HB. 
In this regard, it would be worth exploring the more detailed PT mechanisms with different functionals in future works.
Importantly, the reorganization of HB network before PT might be informative to the fast dynamics captured by the fast-developing infrared spectroscopy.~\cite{09PNAS-Roberts, 11JPCA-Roberts, 15S-Thamer, 18JCP-Napoli, 18JPCB-Carpenter, 19ACS-Yuan}
In spite of this, the validation of presolvation structure with SCAN-based AIMD simulations against spectroscopic experiment awaits future studies.}

\subsection{Dynamic Properties}


\begin{figure}
  \includegraphics[width=8.5cm]{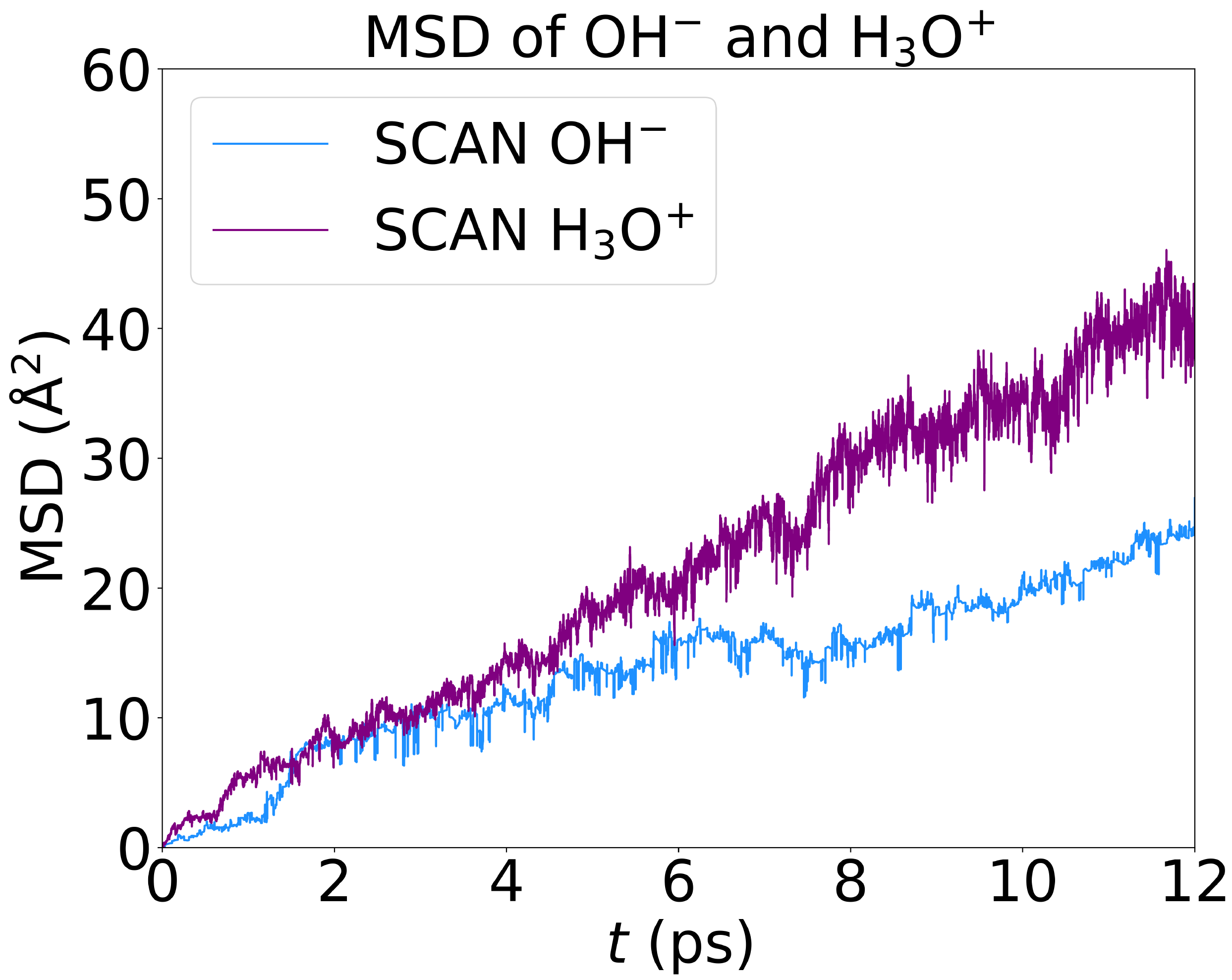}
  \caption{
  (Color online) \MC{Mean square displacements (MSD) extracted from the SCAN trajectories
  of $\rm OH^{-}$(aq) and $\rm H_{3}O^{+}$(aq) are displayed in blue and purple colors, respectively.}
    }  \label{MSD}
\end{figure}

\begin{figure}
  \includegraphics[width=8.5cm]{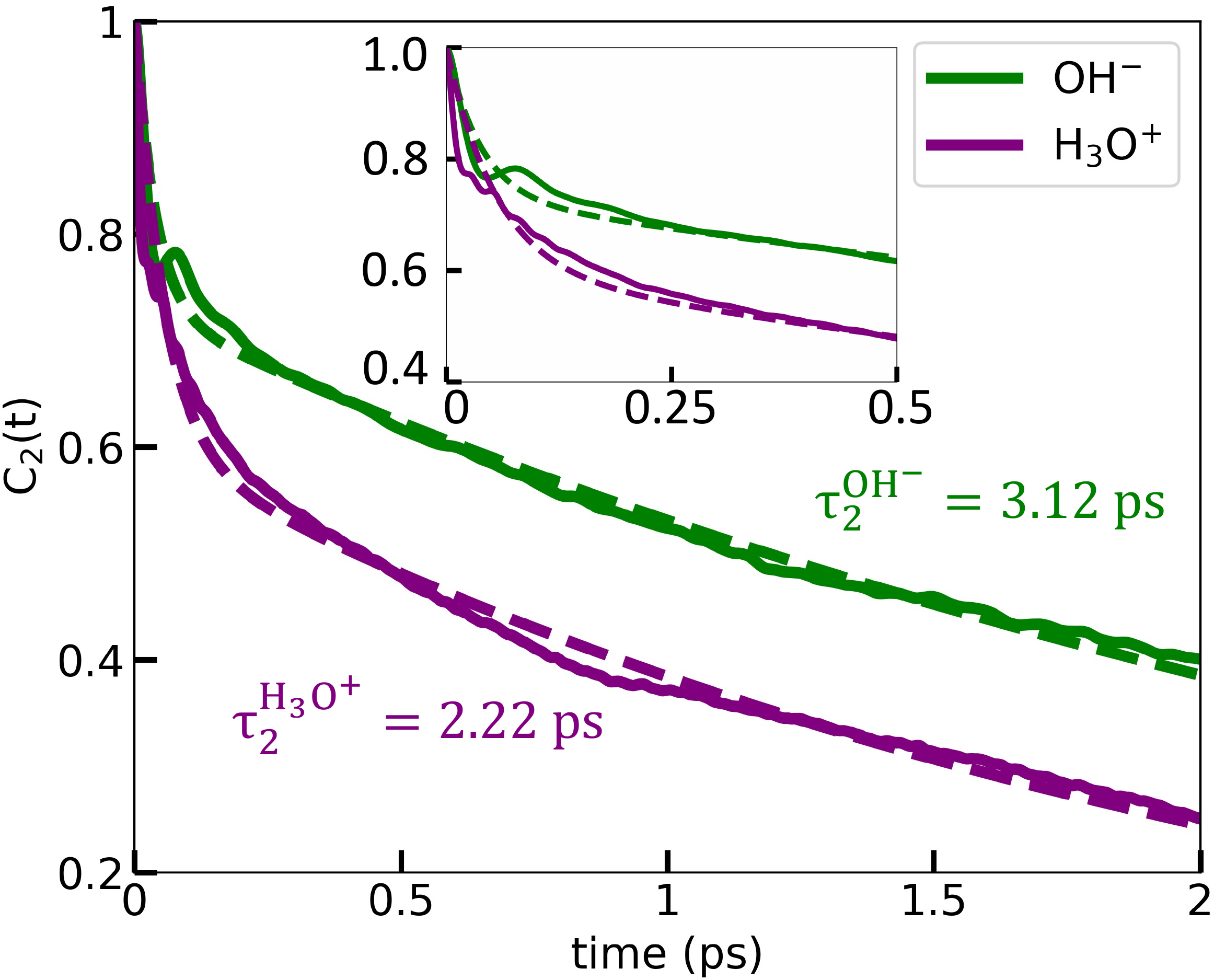}
  \caption{
  (Color online) \RXN{Orientation time correlation function (TCF) $\mathrm{C_{2}(t)}$ of the solvated hydroxide (green) and hydronium (purple) in water. Biexponential fits (dashed lines) in the form of $\mathrm{C(t)=ae^{t/\tau_1}+(1-a) e^{t/\tau_2}}$ for the first 2 ps are also plotted. Here $\mathrm{\tau_1}$ and $\mathrm{\tau_2}$ denote the time scale of the two exponential decays respectively and $\mathrm{a}$ denotes the proportion of the first component. Inset magnifies the TCF and the fitted curves within 0.5 ps.}}  \label{orien}
\end{figure}

\begin{table}
\caption{
Diffusion coefficients of $\rm H_{3}O^{+}$(aq) and $\rm OH^{-}$(aq) ($\rm 10^{-9} m^{-2}/ s$), \MC{$\rm D^{+}$ and $\rm D^{-}$, and the ratio $\rm D^{+}$/$\rm D^{-}$}.
The experimental data~\cite{83RSC-Halle, 90N-Weingartner, 10JPCB-Sluyters} and PBE0-TS results are compared in Ref.~\onlinecite{18NC-Chen} and listed here.
}\label{diffusion}
\begin{tabular}{c|c|c|c|c}
\hline
  & SCAN & PBE0-TS & Exp. ($\rm D_{2}O$) & Exp. ($\rm H_{2}O$) \\\hline 
$\rm OH^{-}$(aq) & 2.9 \MC{$\pm$ 0.7} & 3.7 & 3.2\cite{83RSC-Halle}, 3.1\cite{10JPCB-Sluyters} &  5.4\cite{83RSC-Halle}, 5.2\cite{10JPCB-Sluyters}  \\
\hline
$\rm H_{3}O^{+}(aq)$ & 5.7 \MC{$\pm$ 0.8} & 8.3 & 6.9\cite{83RSC-Halle}, 6.7\cite{10JPCB-Sluyters} & 9.6\cite{83RSC-Halle}, 9.4\cite{10JPCB-Sluyters}   \\
\hline
\MC{$\rm D^{+}$/$\rm D^{-}$} & 1.97 & 2.24 & 2.16\cite{83RSC-Halle}, 2.16\cite{10JPCB-Sluyters} & 1.78\cite{83RSC-Halle}, 1.81\cite{10JPCB-Sluyters}
\\\hline
\end{tabular}
\end{table}

We used the Einstein equation to compute the diffusion coefficient of ions as
\begin{equation}
D = \lim_{t \to \infty} \frac{\| \mathbf{r} (t) - \mathbf{r}(0) \|^{2}}{6t},
\end{equation}
where the mean square displacement (MSD) is defined as $\| \mathbf{r} (t) - \mathbf{r} (0) \|^{2}$
with $\mathbf{r}(t)$ depicts the atom position at time $t$. 
The results are listed in Table~\ref{diffusion}.
\MC{In order to obtain a reasonable value of $D$,
we split the trajectories into multiple segments with a length of 12 ps for each segment and
the initial snapshots were displaced by an interval of 3 ps. Fig.~\ref{MSD} shows the resulting MSDs of solvated hydroxide and hydronium ions, which were taken from these slices of trajectories and averaged.} Here we find that
the diffusion of hydroxide is substantially slower than hydronium,
in agreement with the experimental data and previous analysis.
To be specific, the diffusion coefficients of both water ions from SCAN trajectories
are relatively smaller than those of PBE0-TS.
As listed in Table~\ref{diffusion},
SCAN (PBE0-TS) predicts $D$ of hydronium and hydroxide
to be $\rm 5.7~(8.3)\times10^{-9} m^{2}/s$ and $\rm 2.9~(3.7)\times10^{-9} m^{2}/s$, respectively.
Based on previous analysis of the structural, electronic, and dynamic properties of the two ions,
we infer that the relatively lower $D$ values from SCAN originate 
from the different HB networks described by SCAN and PBE0-TS.
In detail, the oxygen site of hydroxide described by SCAN is more hydrophilic
and the hydroxide anion is more often solvated by a stable hyper-coordinated structure 
with four or even five HBs connected with neighboring water molecules.
The hyper-coordinated structure turns the ion into an asleep mode and hinders the PT events, 
preventing itself from fast structural diffusion. As a result, the PT of hydroxide prefers
the stepwise motion.
On the other hand, the solvated hydronium described by SCAN has a relatively smaller hydrophilicity than PBE0-TS. 
However, the stable three HBs still form between the protons of hydronium and neighboring water molecules.
In this regard, the concerted PT character is preserved for hydronium. 
Notably, a lack of the second neighbors further prevents concerted PTs from occurring,
which substantially lower the frequency of PT bursts in hydronium.
Consequently, the diffusivity of hydronium is lowered as compared to the result from PBE0-TS.
In conclusion, the diffusivities of both hydroxide and hydronium ions are 
lowered from SCAN as compared to those from PBE0-TS, and both functionals predict
that hydroxide diffuses slower than hydronium, in agreement with the experiments.

\RXN{
Furthermore, we calculated the orientation time correlation function (TCF) of the two ions with the SCAN functional, $\mathrm{C_2 (t)= <P_2 (\mathbf{u}(0)\cdot\mathbf{u}(t))>}$, where $\mathrm{\mathbf{u}(t)}$ denotes a unit vector fixed on the ion at time $\mathrm{t}$, $\mathrm{P_2 (x)}$ denotes the second-order Legendre polynomial and $\mathrm{P_2 (x)=(3x^2-1)/2}$.~\cite{11CPL-Ma} 
The unit vector is picked as the vector along the O-H covalent bond for hydroxide,
while it is picked as the normalized sum of unit vectors along all of the three O-H covalent bonds for hydronium. 
The first 2 ps components of the generated TCF are shown in Fig.~\ref{orien}, together with fitted curves of the biexponential function, $\mathrm{C(t)=ae^{t/\tau_1}+(1-a) e^{t/\tau_2}}$, where $\mathrm{\tau_1}$ and $\mathrm{\tau_2}$ denote the time scale of the two exponential decays respectively and $\mathrm{a}$ denotes the proportion of the first component. 
The orientation TCF of both ions exhibit similar biexponentiality as compared to that of water molecule.\cite{11AR-Laage}
The characteristic time of the first component ($\mathrm{\tau_1}$) of both ions are similar in magnitude, indicating that the libration mode of water molecules is largely reserved in the ions. 
However, the two ions yield different relaxation time for the second component ($\mathrm{\tau_2}$). 
In detail,
the TCF of hydronium with a characteristic time $\mathrm{\tau_2}$ = 2.22 ps decays substantially faster than that of hydroxide ($\mathrm{\tau_2}$ = 3.12 ps),
which implies the inactiveness of hydroxide compared to hydronium. 
Therefore, we confirm that the second relaxation process could be primarily due to the reorientation induced by PT as originally proposed by Ma et al.~\cite{11CPL-Ma}
In this way, the difference in the characteristic time $\mathrm{\tau_2}$ between the two ions could be explained by the different PT rates between them.}
%
%

\section{Conclusions}

Accurate AIMD simulations of water with ions including hydronium and hydroxide have long been puzzled and hindered due to the lack of suitable XC functionals.
We proposed that the general-purpose SCAN XC functional could be an appropriate choice in modeling solvated hydronium and hydroxide ions in liquid water. We performed systematic investigations on the 
structural, electronic{\XY{,}} and dynamical properties of both solvated hydroxide and hydronium ions.
Specifically, we studied the amphiphilic properties, the solvation structures, the electronic structures, 
and the dynamic properties of the two ions.
Although the hybrid functional plus van der Waals correction (PBE0-TS) provides a satisfactory 
picture for the two ions, the computational costs are extremely high.
In this regard, we compared the SCAN results to the counterparts 
from PBE0-TS and the main conclusions are summarized as follows.

The SCAN functional qualitatively 
reproduces the amphiphilic nature of both water ions 
as compared to the PBE0-TS functional. 
In detail, the oxygen site of hydroxide prefers to accept four or even five HBs
while donating an unstable H-bond, 
and SCAN functional predicted substantially more hydroxide ions that accept 5 HBs than PBE0-TS.  
Meanwhile, hydronium stably donates all of its three protons to neighboring molecules and hardly accepts HBs.
Due to the increased number of accepted HBs at the oxygen site of hydroxide, 
SCAN produces a pot-like solvation structure of hydroxide,
which generally agrees well with the result from PBE0-TS and experiment.
%
Note that the different features of the HB network and solvation structure of hydroxide and hydronium
originate from their electronic structure. Thus, we analyzed
the distributions of MLWF centers, and found 
the locations of the lone pair and bonding pair electrons 
respectively determine the hydrophobic and hydrophilic environments 
of $\rm OH^{-}$(aq), $\rm H_2O$, and $\rm H_3O^{+}$(aq) molecules.
Generally, compared to the results from PBE0-TS, we found SCAN yielded
a closer distance between the lone pairs and the oxygen atom and a larger distance between the bonding pairs and the protons.
As a result, the hydroxide anion from SCAN becomes more hydrophilic on its both ends while
the hydronium cation from SCAN becomes slightly more hydrophobic on its two ends.
The two competing effects affect the HB network described by SCAN and subsequently 
determine the dynamic property, particularly for the ubiquitous PT events. \MC{In addition, we analyzed the dipole moments of the neighboring water molecules of the two ions and found both exchange-correlation functionals yield similar values.}
On \MC{the} one hand, the hyper-coordinated structure turns the hydroxide into an asleep mode 
and hinders the PT events; both SCAN and PBE0-TS predict that the PT of hydroxide prefers
the stepwise motion.
On the other hand, the concerted PT character is preserved for hydronium.
However, a lack of the second neighbors further prevents concerted PTs from occurring,
significantly lowering the frequency of PT bursts in hydronium.
Consequently, the diffusivity of hydronium is lowered as compared to the result from PBE0-TS.
\RX{
Besides, we also reported two characteristic non-diffusional processes, i.e., trap and revisit,
indicating that PT classification is non-trivial and further studies are needed.}
In summary, the diffusivities of both hydroxide and hydronium ions are
lowered from SCAN as compared to those from PBE0-TS, and both functionals predict
that hydroxide diffuses slower than hydronium, in agreement with the experiments.



\MC{The affinity of the two ions to the water-air surface is another interesting topic for both experiments~\cite{12PNAS-Mishra, 08CPL-Poul} and simulations.~\cite{09CPL-Mundy, 14JPCB-Baer,15JACS-Tse}
For example, no consensus has been reached on the interfacial preferences of the two water ions.~\cite{07PNAS-Buch, 08JACS-Kudin, 08CPL-Pegram, 09PCCP-Gray, 09FD-Beattie}
Furthermore, it is still an unsettled issue about the propensity of hydrated hydroxide at the hydrophobic interface.~\cite{20L-Yang, 10COCIS-Ralf, 15RSC-Fang, 09PNAS-Tian}
Typically, modeling the water-air interface or the hydrophobic material-water interface from first-principles methods is computationally demanding~\cite{11N-Stiopkin} because the atomic model includes both bulk and interfacial atoms.~\cite{16CR-Agmon}
In this work, we demonstrate that AIMD simulations with the SCAN functional may shed new lights on the above issues. This is because the SCAN functional is a general-purposed exchange-correlation functional, which provides an excellent choice in predicting the structural and dynamic properties of both hydronium and hydroxide ions in the bulk water.
}

\RXN{The coupling of nuclear quantum effects (NQEs) and the SCAN XC functional is also an important issue, since NQEs have significant impacts on the electronic and HB structure of water.~\cite{21B-Tang}
For instance, previous studies have shown that strong HBs of water molecules are strengthened by NQEs while weak ones are weakened.~\cite{16CR-Ceriotti, 11PNAS-Li, 13PNAS-Ceriotti, 09JCP-Habershon}
The two competing effects coexist in water and generally show a weakened effect on the HBs.~\cite{16JCP-Marsalek, 16JCTC-Gasparotto, 21JACS-Calio}
Based on the pioneering works by Tuckerman et al.,~\cite{02N-Tuckerman, 99N-Marx, 00JPCM-Marx} one could expect that the energy barrier of PT events in water would be lowered by NQEs, which may further affect the structural diffusion of water ions. 
However, the influences of NQEs on HB network also rely on the underlying XC functional, so how the SCAN functional and NQEs couple and its influences on PT still requires future in-depth studies.}

In conclusion, the SCAN XC functional that adopted in AIMD simulations
is already known as an excellent XC functional 
in describing a variety properties of liquid water. In this work, 
we further demonstrate that SCAN stands out as
the long awaited general-purpose XC functional that can 
accurately and efficiently predict various properties 
of solvated hydroxide and hydronium ions in ambient liquid water.
\RX{In addition, the SCAN functional opens up several interesting topics that worth exploring in future. For example, the origin of the water dipole moments surrounding the water ions, the classification of rattling events and the statistical description of the trapping and revisit phenomena of PT events, the spectroscopy of the solvated hydronium and hydroxide ions, and the nuclear quantum effects in affecting the physical properties of ions.}
We are looking forward to the usage of SCAN in modeling 
a variety of important physical and chemical processes that involve 
hydroxide and hydronium ions. 
%


$\\$
{\bf Acknowledgements}
$\\$
The work of M.C. is supported by the National Science Foundation of China under Grant No. 12122401 and No. 12074007. Parts of the numerical simulations were performed on the High Performance Computing Platform of CAPT. The work by C.Z. and X.W. was supported by National Science Foundation through Award No. DMR-2053195
and No. DMR-1552287. This research used resources of the National Energy Research
Scientific Computing Center (NERSC), which is supported by the U.S. Department of Energy
(DOE), Office of Science under Contract No. DE-AC02-05CH11231.

$\\$
{\bf Availability of Data}
$\\$
The data that support the findings of this study are available from the corresponding author upon reasonable request.

\bibliography{reference}

\end{document}